\documentclass[aps,amsmath,amssymb,twocolumn,showpacs]{revtex4-1}

\usepackage{graphicx}
\usepackage{dcolumn}
\usepackage{bm}
\usepackage{amsmath}
\usepackage{amssymb}
\usepackage{color}
\usepackage{bbold}
\usepackage{bm}
\usepackage{longtable}
\usepackage{varioref}
\usepackage{mathtools}
\usepackage{tabls}
\usepackage{array,url,kantlipsum}

\newcommand{\cl}[1]{\hat{\mathcal{#1}}}

\newcommand{\ket}[1]{|#1\rangle}
\newcommand{\bra}[1]{\langle#1|}
\newcommand{\ave}[1]{\langle #1 \rangle}

\newcommand{\tr}{\text{Tr}}
\newcommand{\comm}[1]{{\color{black}#1}}

\newcommand{\llangle}[1][]{\savebox{\@brx}{\(\m@th{#1\langle}\)}%
  \mathopen{\copy\@brx\kern-0.5\wd\@brx\usebox{\@brx}}}
\newcommand{\rrangle}[1][]{\savebox{\@brx}{\(\m@th{#1\rangle}\)}%
  \mathclose{\copy\@brx\kern-0.5\wd\@brx\usebox{\@brx}}}
	
\begin{document}
\title{Perturbative approach to weakly driven many-particle systems in the presence of approximate conservation laws}
\author{Zala Lenar\v{c}i\v{c}}
\author{Florian Lange}
\author{Achim Rosch}
\affiliation{Institute for Theoretical Physics, University of Cologne, D-50937 Cologne, Germany}

\begin{abstract}
We develop a Liouville perturbation theory for weakly driven and weakly open quantum systems in situations when the unperturbed system has a number of conservations laws. If the perturbation violates the conservation laws, it drives the system to a new steady state which can be approximately but efficiently described by a (generalized) Gibbs ensemble characterized by one Lagrange parameter for each conservation law. The value of those has to be determined from rate equations for conserved quantities. Remarkably, even weak perturbations can lead to large responses of conserved quantities. We present a perturbative expansion of the steady state density matrix; first we give the condition that fixes the zeroth order expression (Lagrange parameters) and then determine the higher order corrections via projections of the Liouvillian. The formalism can be applied to a wide range of problems including two-temperature models for electron-phonon systems, Bose condensates of excitons or photons or weakly perturbed integrable models. We test our formalism by studying interacting fermions coupled to non-thermal reservoirs, approximately described by a Boltzmann equation.
\end{abstract}

\pacs{
05.70.Ln, 
05.30.-d,	
03.65.Yz,	
02.30.Ik, 
}

\maketitle

In equilibrium many particle systems can be efficiently described by Gibbs ensembles, characterized by one Lagrange parameter (inverse temperature, chemical potential) for each exactly conserved quantity (energy, particle number).  Such a simple but powerful description is, in general, absent for driven non-equilibrium systems. As long as weak perturbations, which drive the system out of equilibrium, have small effects one can resort to perturbation theory. Kubo formulas, for example, describe the response to time-dependent Hamiltonian perturbations. Also for open systems described by Lindblad operators similar formulations of perturbation theory can be developed \cite{benatti11,li14,li16,maizelis14,garcia09,kessler12,znidaric15,Reiter12,lemos17}. If the density matrix of the unperturbed system is not unique  (as is the case for all Hamiltonian systems)
one has to use degenerate Liouville perturbation theory \cite{benatti11,znidaric15}.

There are many situations where even weak perturbations of interacting many-particle systems can have large effects. A famous example is the Bose-Einstein condensation of excitons and polaritons \cite{butov01,kasprzak06,balili07}: irradiation by light creates more and more excitons which equilibrate approximately and form a Bose-Einstein condensate. Importantly, the exciton number is approximately conserved thus even weak pumping can compensate for exciton losses and leads to large densities of these excitations. Similarly, condensations of photons or magnons have been observed \cite{klaers10,klaers12,schmitt14,demokritov06,chumak09}. 


This is a general phenomenon: whenever approximate conservation laws exist, small perturbations which weakly break those conservation laws can drive the system out of equilibrium, to a steady state completely different from the initial one. \comm{In the example given above, the approximately conserved quantity is the exciton number. Other examples use the approximate conservation of spin to induce large nuclear spin polarization \cite{walker97} for medical applications or use the weak coupling of electrons and phonons to induce vastly different temperatures in the two subsystems \cite{anisimov74,armstrong81,eesley86,allen87,singh10}. 
In all cases, the  long-time steady state -- a Bose-Einstein condensate for the exciton example -- is  very different from any thermal state. }Nevertheless, simple approximate theoretical descriptions can be found, e.g., by introducing an effective chemical potential for the excitons.

As we have argued recently in Ref.~\cite{lange17}, this approach works quite generally: 
generalized Gibbs ensembles characterized by Lagrange parameters for {\em approximately} conserved quantities
give an approximate description of the steady state of weakly perturbed driven systems. They describe situations where the driven system is completely different from the initial one, with Lagrange parameters determined by the perturbations. \comm{The nearly conserved quantities -- e.g. the exciton number in the example discussed above -- show a strong (non-linear) response to 
weak perturbations.}
In this paper we develop a perturbation theory around these generalized Gibbs ensembles and show that deviations remain small as long as the perturbations driving the system out of equilibrium are small. This is, however, only true if the correct zeroth order approximation, i.e. the correct generalized Gibbs ensemble, is chosen as the reference point of the perturbative expansion.
The formalism is developed both for Lindblad and unitary time-periodic (Floquet) perturbations. 

From this perspective our investigation is related to other Floquet studies in the presence of interactions and/or coupling to the environment \cite{dehghani14,shirai15,iadecola15,liu15,iwahori16,shirai16,seetharam15,iwahori17,genske15}.
For example, in Ref.~\cite{genske15} experiments of the Esslinger group on a Floquet-realization of the Haldane model \cite{esslinger} have been modeled based on the derivation of a Floquet-Boltzmann equation \cite{genske15,seetharam15}. The concept of a (time-dependent) temperature was used to describe the heating of interacting Floquet systems for situations where energy-conserving scattering processes dominate scattering events, where the energy changes. 
A related approach is taken when describing weakly-coupled electron-phonon systems out of equilibrium. Here one often associated two different temperatures to the two subsystems with energies that are separately approximately conserved \cite{anisimov74,armstrong81,eesley86,allen87,singh10,sentef13}. 
In Ref.~\cite{lange17}
we used generalized Gibbs ensembles to describe the steady state of  spin-chains coupled to phonons and driven by an external laser, modeled by a periodic driving. In these systems an infinite number of approximately conserved quantities exists due to the integrability of the underlying spin-1/2 Heisenberg model.
As the heat current is also approximately conserved, one can realize large heat currents by weak driving.

\comm{From a more general point of view, our theoretical approach is based on the idea to project the dynamics of the density matrix on a few relevant degrees of freedom -- in our case the approximate conservation laws. Such projection formalisms are widely used by many authors \cite{Robertson66, Robertson68,kawasaki73,Ochiai73,Willis74,Grabert74,Koide02,breuer02}. In contrast to these studies, we will use generalized Gibbs ensembles as a reference point and we focus our calculation on the computation 
of  the steady state.}

The article is organized in the following way: Sec.~\ref{SecModel} prepares the perturbative setup, Sec.~\ref{SecLeading} gives the prescription on how to find the zeroth order approximation to the steady state via the introduction of a super-projector. The latter is used in Sec.~\ref{SecPert} to determine corrections around the approximate description of the steady state both for Lindblad and unitary time-periodic (Floquet) perturbations.
The applicability of our degenerate perturbation theory is tested in Sec.~\ref{example} on an example of \comm{interacting fermions coupled to non-thermal reservoirs which induce weak gain and loss of particles. We first perform an exact calculation on small system and then use approximate description by Boltzmann dynamics in the thermodynamic limit.}

\section{Model}\label{SecModel}
We consider a many-particle quantum system described by 
the density operator $\rho$. Its dynamics is determined from the Liouville equation $\dot{\rho}=\cl{L} \rho$, where   $\cl{L}$ is the Liouvillian super-operator. We consider situations where a Hamiltonian system is weakly perturbed. Therefore we split $
\cl{L}=\cl{L}_0 + \cl{L}_1$ into two parts, where $\cl{L}_0$ describes the dominant unitary Hamiltonian evolution and $\cl{L}_1$ a weak perturbation of strength $\epsilon$ which drives the system out of equilibrium,
\begin{align}
\cl{L}_0\rho=-i[H_0,\rho],\quad
\cl{L}_1\rho = \left\{ \begin{array}{ll}
-i[\epsilon H_1,\rho],\\
\epsilon\cl{D}\rho.
\end{array} \right.
\end{align} 
Most importantly, the unperturbed system described by $H_0$ has the property of having additional conserved quantities $C_i$, $[H_0,C_i]=0$, $i=1,..,N_{\textrm c}$. We are interested both in cases where $N_{\textrm c}$ is small or when $N_c$ is extensive as is the case when $H_0$ is integrable.
The perturbation $\cl{L}_1$ is of the unitary or/and of the Markovian Lindblad form. In $H_1$ we consider both static and time-dependent perturbations which are periodic in time. The Lindblad dynamics is described by  
\begin{equation}\label{lindblad}
\cl{D}\rho=\sum_{\alpha}\big(L_\alpha \rho L_\alpha^{\dagger} - \frac{1}{2}\{L_\alpha^{\dagger} L_\alpha,\rho\}\big)
\end{equation}
with Lindblad operators $L_\alpha$. We consider only translational invariant perturbations and situations where a unique (Floquet-) steady state is obtained. 
We assume that the perturbations break all conservation laws considered above. Remaining exactly conserved quantities are fixed by initial conditions and can easily be included in the theoretical description but are omitted in the following to simplify notations.

We are mainly interested in the nonequilibrium steady state. We split its  density operator $\rho_\infty$ into the zeroth order approximation $\rho_0$  and corrections $\delta\rho$,
\begin{equation}\label{rhoLong}
\lim_{t\to\infty}\rho(t)=\rho_{\infty}=\rho_0 + \delta\rho.
\end{equation}
with  $\rho_0=\lim_{\epsilon\to 0}\lim_{t\to \infty} \rho(t)$. Note that the limits $t\to \infty$ and $\epsilon \to 0$ do not commute: as we will show, small perturbations can completely change the density matrix in the long-time limit.

For perturbations periodic in time, the density matrix is periodically oscillating in the long time limit but one can still use the above formulas by interpreting them in Floquet space, see Section \ref{SecPeriodic0App}.
The higher order corrections $\delta\rho$ are formally given by
\begin{equation}\label{EqDeltaRho}
\delta\rho=- \cl{L}^{-1}\cl{L}_1\rho_0, \quad
\cl{L}^{-1} \to \lim_{\eta\to 0}(\cl{L}-\eta\hat{1})^{-1}
.
\end{equation}
where we used $\cl{L}\rho_\infty=\cl{L}_1\rho_0 + \cl{L}\delta\rho=0$ and $\cl{L}_0\rho_0=0$. The inverse $\cl{L}^{-1}$ should be interpreted using infinitesimal regularization $\eta$, see App.~\ref{AppRegularize}. Due to the  conservation laws of $H_0$, $\cl{L}_0\rho_0=0$ has no unique solution. While Eq.~\eqref{EqDeltaRho} is formally valid for arbitrary $\rho_0$ with $\cl{L}_0\rho_0=0$, the correction $\delta\rho$ will only be small if  $\rho_0$ is correctly chosen as discussed in the next section.

\section{Zeroth order: generalized Gibbs ensemble}\label{SecLeading}
We have defined $\rho_0$ to fulfill the equation  $\cl{L}_0\rho_0=0$. In the thermodynamic limit, generic steady states of interacting many-particle system with (quasi-local) conservation laws $C_i$ approach states which can be described by a (generalized) Gibbs ensemble (GGE)
\begin{equation}\label{EqGGE}
\rho_0
=\frac{e^{-\sum_i \lambda_i C_i}}{\tr{[e^{-\sum_i \lambda_i C_i}]}},
\end{equation} 
More precisely, the system as a whole may be in a different ensemble (e.g. a canonical ensemble) but for the computation of local observables one expects that the different ensembles are generically equivalent in the thermodynamic limit. One manifestation of this is the eigenstate thermalization hypothesis \cite{polkovnikov11,nandkishore15} which argues that a generic pure state becomes equivalent to a Gibbs state in the thermodynamic limit. Eq.~\eqref{EqGGE} has also been extensively tested for integrable models
\cite{rigol07,calabrese07gge,cramer08gge,barthel08gge,essler12gge,pozsgay13gge,fagotti13gge,fagotti13gge2,fagotti14gge,fagotti14agge,wouters14gge,pozsgay14gge,sotiriadis14gge,goldstein14gge,brockmann14gge,pozsgay14agge,rigol14gge,mestyan15gge,ilievski15agge,eisert15gge,ilievski16gge,piroli16gge,calabrese11gge,calabrese12gge,calabrese12agge,fagotti13gge,caux13gge,essler16gge,essler17gge,ilievski16}
which appear to approach a state described by $\rho_0$  after a quench. For certain local initial states (satisfying cluster decomposition property) this has been shown rigorously \cite{ilievski15agge}. It has been argued recently that there exist a protocol based on truncated GGEs, convergent in number of included conservation laws \cite{pozsgay17}.

Note that the parameters $\lambda_i$ in Eq.~\eqref{EqGGE} have not yet been determined.
Most importantly, they are {\em not} fixed by initial conditions as the $C_i$ are not conserved in the presence of the perturbations described by $\cl{L}_1$. Instead the $\lambda_i$ or, equivalently, the expectation values of the $C_i$ have to be determined from rate equations governed by the weak perturbation{}s. As we will show next, these will lead to changes of the $\lambda_i$ of order $\mathcal{O}(1)$.

\subsection{Determination of $\lambda_i$}\label{SecCondGGE}
All GGEs satisfy $\cl{L}_{0}\rho_0=0$ by definition. Therefore the perturbation $\cl{L}_1$  fixes the steady state parameters $\{\lambda_i\}$. Technically, we determine the $\{\lambda_i\}$ from the condition that the state is stationary in combination with the condition that $\delta \rho$ should be small in the limit $\epsilon \to 0$. 
The first condition ensures that $\ave{\dot{C}_i}=0$, which is  evaluated using straightforward perturbation theory
\begin{align}
\ave{\dot{C}_i}
&=\tr[C_i \cl{L} \rho_\infty]\nonumber \\
&=\tr[C_i \cl{L}_1 \rho_0] + \tr[C_i \cl{L}_1 \delta\rho]\approx \tr[C_i \cl{L}_1 \rho_0]  \label{EqCdot}
\end{align}
for $i=1,...,N_{\textrm c}$.
Above we used that $\tr[C_i \cl{L}_0 \delta\rho]=0$ due to 
\comm{$\cl{L}_0^\dagger C{}_i =i[H_0,C_i]=0$ and that 
$\delta \rho$ is small. Note that the adjoined of a Liouvillian is defined by the equation
$\tr[A \cl{L} \rho]=\tr[ ( \cl{L}^\dagger A) \rho]$.}
 We therefore obtain from $\ave{\dot{C}_i}=0$ the condition fixing $\rho_0$
\begin{align}
 \tr[C_i \cl{L}_1 \rho_0] \stackrel{!}{=} 0  \label{EqRho0}
\end{align}
If $\tr[C_i \cl{L}_1 \rho_0]=0$ is trivially fulfilled for all $\rho_0$, one has to consider higher-order perturbation theory, see below.

Assuming for the moment that $\tr[C_i \cl{L}_1 \rho_0]\neq 0$ for all $i=1,...,N_{\textrm c}$ and generic $\rho_0$, then the $N_{\textrm c}$ equations (\ref{EqRho0}) can be used to fix the $N_{\textrm c}$ Lagrange parameters $\lambda_i$.
These equations can be viewed as  {\it rate equations}, which describe how the perturbations change the approximately conserved $C_i$. Within the approximation used above, the size of the perturbation, i.e. the value of $\epsilon$, completely drops out of the equation for the steady state. Therefore it fixes the Lagrange parameters (and, accordingly, also $\rho_0$)
to order $\epsilon^0$. While to this order the size of the perturbation is not important, its structure determines the Lagrange parameters and induces changes of order $1$.

We now consider situations where $\tr[C_i \cl{L}_1 \rho_0]=0$ for all $\rho_0$. In this case one has to 
use perturbation theory at least to order $\epsilon^2$ to fix $\rho_0$ to order $\epsilon^0$.
This is always the case for Hamiltonian perturbations where $\cl{L}_1 \rho_0=-i  [\epsilon H_1,\rho_0]$ (this follows from $[C_i,\rho_0]=0$ and the cyclic property of the trace). It corresponds
to the well-known fact that transition rates, e.g., calculated from Fermi's golden rule, are always quadratic in the perturbation. To calculate $\ave{\dot{C}_i}$ to order $\epsilon^2$, $\delta\rho$ has to be expanded up to $\mathcal{O}(\epsilon)$ using Eq.~\eqref{EqDeltaRho} and the relation
\begin{equation}\label{EqXY}
(X+Y)^{-1}=X^{-1}-(X+Y)^{-1}YX^{-1},
\end{equation}
with $X=\cl{L}_0$ and $Y=\cl{L}_1$. 
All inverses here and in the following are regularized as in Eq.~(\ref{EqDeltaRho}).
From this equation we find that
$\delta\rho=-\cl{L}_0^{-1} \cl{L}_1\rho_0 + \cl{L}^{-1}\cl{L}_1\cl{L}_0^{-1}\cl{L}_1\rho_0$.
A priori, it is not obvious that the first term linear in $\cl{L}_1$ in this expression is large compared to the second one quadratic in $\cl{L}_1$
due to possible singularities in  $\cl{L}^{-1}$ and we will indeed show that this is in general {\em not} the case. Nevertheless, using the machinery developed in Sec.~\ref{SecPert}, we can show that for the rate equation under discussion, one can make this approximation, see App. \ref{AppUniGGE}. Therefore we obtain from 
$\tr [C_i \cl{L}_1\delta\rho]=0$
the condition 
\begin{equation}\label{EqUniPert}
\tr[C_i \cl{L}_1 \cl{L}_0^{-1} \cl{L}_1\rho_0]\stackrel{!}{=}0.
\end{equation} 
As above, the size of $\epsilon$ drops out of this equation which therefore fixes 
 $\rho_0$ to order $\epsilon^0$ in situations where (\ref{EqRho0}) is trivially fulfilled.
For combined unitary and Markovian perturbations terms from Eq.~(\ref{EqRho0}) and Eq.~(\ref{EqUniPert}) have to be considered simultaneously.

There can be situations where there is no contribution of order $\epsilon^2$ to the decay rate of a conservation law such that the left-hand side of both Eq.~\eqref{EqRho0} and Eq.~\eqref{EqUniPert} vanishes. This happens, for example, for an integrable Heisenberg chain perturbed by next-nearest neighbor interactions as discussed in Ref.~\cite{jung06} many years ago. 
Also in Refs.~\cite{lenarcic15,benenti09} where perturbed $L$-site Ising model and Heisenberg model with strong magnetic field were studied, the boundary Lindblad operators fix the density matrix completely only after terms up to order $L$ or $L-1$ were included, for odd and even system sizes, respectively.
In such cases, one has to use Eq.~\eqref{EqXY} recursively to obtain higher-order corrections.

A non-trivial test of the order of perturbation that fixes $\rho_0$ can be obtained by exact (numerical) calculation of the Liouville gap, 
 $\Lambda\equiv \min(|Re \ \lambda|; Re \ \lambda<0), \ \cl{L} \rho=\lambda\rho$, on a system of smaller size. If scaling $\Lambda \sim \epsilon^k$ is obtained then $\rho_0$ is determined by the condition of order $\epsilon^k$, which is for example $k=1$ for Eq.~(\ref{EqRho0}) and $k=2$ for Eq.~(\ref{EqUniPert}).

\subsection{Periodic  driving}\label{SecPeriodic0App}
Above we have discussed stationary states obtained for $t\to \infty$ in systems with time-independent Liouvillians. The same approach can, however, also be used for systems with periodic driving. Here we focus on a case where $\cl{L}_0$ and the associated conservation laws $C_i$ are time-independent, while $\cl{L}_1(t)=\cl{L}_1(t+T)$ with period $T$. In this case only minor modifications of the formulation given above are necessary.  
In the long-time limit the density matrix always has oscillatory (Floquet) components, 
\begin{equation}\label{EqFloquet}
\rho(t)=\sum_n e^{-in\omega t} \rho^{(n)}, \ n\in\mathbb{Z},	
\end{equation} 
where $\rho^{(-n)}=\rho^{(n)\dagger}$, $\omega=2\pi/T$. The analog of the unique stationary state of Eq.~\eqref{rhoLong} is  a state with time-independent Floquet components $\rho^{(n)}$.
Due to the limit $\epsilon\to 0$ the zeroth order $\rho_0$ contains only time independent $\rho^{(n=0)}$ component.
It is useful to organize the Floquet components into a vector $\boldsymbol \rho=\{\dots \rho^{(-1)},\rho^{(0)}, \rho^{(1)},\dots\} $ and promote the Liouvillian into a (static) matrix 
\begin{align}
\boldsymbol{\cl{L}}&=\boldsymbol{\cl{L}}_0+\boldsymbol{\cl{L}}_1 \\
\boldsymbol{\cl{L}}_1^{nm}&=\cl{L}^{n-m}_1=\frac{1}{T} \hspace{-.1cm} \int_0^T \hspace{-.2cm} \cl{L}_1(t) e^{i \omega (n-m) t} dt  \\
\boldsymbol{\cl{L}}^{nm}_0&=(i n \omega+\cl{L}_0) \delta_{nm}
\end{align}
Then the condition for $\rho_0$, Eq.~(\ref{EqUniPert}), has to be reformulated in the above sense: with $\rho_0$ and $C_i$ interpreted as vectors with non-zero $n=0$ component and $\boldsymbol{\cl{L}_1}$, $\boldsymbol{\cl{L}_{0}}$ as matrices. $\boldsymbol{\cl{L}_1}$ contains off-diagonal terms due to periodic driving and diagonal terms in case of static perturbations. $\boldsymbol{\cl{L}_{0}}$ contains in addition to diagonal terms due to $\cl{L}_0$ also $in\omega \delta_{nm}$ due to explicit time dependence of $\rho(t)$, 
$\dot\rho=\sum_n e^{-in\omega t}(-in\omega \rho^{(n)} + \dot\rho^{(n)})$. Stationarity of conserved quantities $C_i$, Eq.~(\ref{EqUniPert}), in case of periodic driving applies to their time averages.

\subsection{Projection operators and effective forces}\label{SecForces}
When investigating corrections  $\delta \rho$ of the steady state to the zeroth order approximation $\rho_0$ (or when investigating the time-dependence of $\rho(t)$), 
one has to distinguish corrections {\em within} the slow subspace with $\cl{L}_0 \delta \rho_\|=0$ from those perpendicular to this subspace with  $\cl{L}_0 \delta \rho_\perp\neq 0$. Therefore it is useful to introduce a super-projector $\hat{P}_{\rho_0}$ which projects on the (density-matrix) space tangential to the GGE manifold at the expansion point $\rho_0$. In the following, we will omit the argument $\rho_0$, using $\hat{P}=\hat{P}_{\rho_0}$ to avoid a cluttering of notations.

Small changes  within the manifold around $\rho_0$ can be parametrized by  $\delta \rho_\|=\sum_i \delta \lambda_i \, \partial\rho_0/\partial \lambda_i$ where $\delta \lambda_i$ are arbitrary infinitesimal changes of the Lagrange parameters. The super-operator $\hat P$, projecting on these density matrices, and its complement $\hat Q$  are uniquely given by \comm{ \cite{Robertson68}}
\begin{align}\label{EqP}
\hat{P} X &\equiv -\sum_{i,j} \frac{\partial \rho_0}{\partial \lambda_i} \ (\chi^{-1})_{ij} \tr[C_j X] \\
\hat{Q} X &\equiv (\hat 1 - \hat P) X =X-\hat P X
\end{align}
where 
$\chi_{ij}=-\tr[C_i \, \partial\rho_0/\partial\lambda_j]
=\ave{C_i C_j}_{_{0,\textrm c}}$ is a matrix of generalized susceptibilities. 
We use the notation \comm{$\tr[A\rho]=\ave{A}$} and $\tr[A\rho_0]=\ave{A}_{_0}$ for expectation values with respect to $\rho_0$. In addition, $\ave{AB}_{_{0,c}}=\ave{AB}_{_0}-\ave{A}_{_0}\ave{B}_{_0}$ stands for connected correlation function.
$\hat{Q}$ and $\hat P$ have the property $\hat{Q}^2=\hat{Q}$ and  $\hat{P}^2=\hat{P}$  with $\hat{P} \delta \rho_{\|}=\delta \rho_{\|}$ and are therefore projectors. By construction, $\hat{P} \delta \rho$  is for arbitrary $\delta \rho$ a linear combination of $ \partial\rho_0/\partial \lambda_i$ and therefore an element of the tangential space. Note that $\hat P \rho_0\neq \rho_0$ as $\hat P$ is a not a projector on the space of GGE density matrices but instead a projector on the tangential space at $\rho_0$. 
\comm{Note that Ref.~\cite{kawasaki73} uses an alternative projector,
which adds an extra term to guarantee $\hat P \rho_0=\rho_0$. 
}

The super-operator $\hat P^\dagger$, adjoint to $\hat P$, has also a direct physical interpretation.
The natural scalar product within our approach is  $\tr[ A \delta \rho]$, where $A$ is an operator and $\delta \rho$ a density matrix, and therefore  $\tr [(\hat P^\dagger A) \delta \rho]=\tr [A (\hat P \delta \rho)]$. We obtain
\begin{equation}\label{EqPAdj}
\hat{P}^\dagger A
= -\sum_{ij} C_i \ (\chi^{-1})_{ji} \ \tr\left[A \frac{\partial \rho_0}{\partial \lambda_j}\right].
\end{equation}
$P^\dagger$ acts on operators and maps each operator onto the space of conserved operators.
It gives the part of an operator $A$ which does not decay when the dynamics of the system is described by $\cl{L}_0$ only. The super-operator $\hat P^\dagger$ naturally shows up 
when studying the dynamics of systems with conservation laws $C_i$. For example, the projection operator used in the memory matrix formalism \cite{forster75,mori65,zwanzig60,jung06,jung07a} can (in this case) be identified with $\hat P^\dagger$. The operator can also be used to express the Drude weight of conductivities using the seminal results of Mazur \cite{mazur69} and Suzuki \cite{suzuki71}. The Drude weight $D(T)$ is defined as the prefactor of a $\delta$-function in the optical conductivity, 
$\text{Re}[\sigma(\omega)]= \pi D(T) \delta(\omega) + \sigma_{\rm reg}(\omega)$. 
At finite temperatures $T$, the Drude weight is finite in situations where conservation laws $C_i$ prohibit the decay of the current. Therefore the Drude weight in a thermal state \cite{mazur69,suzuki71,zotos97}  can simply be written in terms of the static cross susceptibility of $J$ and $\hat P^\dagger J$,
$D(T)=\frac{\beta}{L} \langle (\hat P^{\dagger} J) J \rangle_c$ where $\beta=1/T$ and $L$ is the system size.

Below, we will heavily rely on  $\hat P$ when deriving perturbation theory for the stationary state. $\hat{P}$ can, however, also be used to define generalized forces allowing to track the changes of the Lagrange parameters
during time evolution \comm{\cite{Robertson66}}. This allows to calculate, e.g., the heating of a driven system, and thus to obtain an intuitive picture on the dynamics. Assuming that a state $\rho_0(t)$ with time-dependent Lagrange parameters $\lambda_i(t)$ describes the system approximately,  we can use
\begin{align}\label{EqForce}
\hat{P}\dot{\rho}&\approx  \sum_i \frac{\partial \rho_0}{\partial \lambda_i} \frac{\partial \lambda_i}{\partial t} =  \sum_i \frac{\partial \rho_0}{\partial \lambda_i}  F_i. 	\\
\dot{\lambda}_i &= F_i \approx- \sum_j (\chi^{-1})_{ij} \tr[C_j \dot \rho ]= -\sum_j (\chi^{-1})_{ij} \ave{\dot C_j} \notag
\end{align}
to obtain generalized forces $F_i$ governing to leading order the dynamics of the Lagrange parameters. 
Depending on the studied case it is enough to include only the dominant contribution, e.g., $ F_i=- \sum_j (\chi^{-1})_{ij} \tr[C_j \cl{L}_1 \rho_0]$.
In this paper, we will mainly focus on corrections to the stationary state, but we will discuss dynamics briefly in the context of the Boltzmann equation.

The stationarity condition (\ref{EqRho0}) can be then rewritten as 
\begin{align}\label{rho0Pro1}
 \hat{P}(\cl{L}_1\rho_0)=0.
 \end{align} 
 Geometrically it means that $\cl{L}_1\rho_0$ must be perpendicular ($\perp$) to the slow manifold. Similarly the condition for unitary perturbation, Eq.~(\ref{EqUniPert}), transcripts into 
 \begin{align}\label{rho0Pro2}
\hat{P}(\cl{L}_1 \cl{L}_0^{-1}\cl{L}_1\rho_0)=0.
\end{align}
 In both cases we require that the  forces driving the system have effectively to be perpendicular to the GGE manifold.

Note also that in case of periodic driving $\hat{P}$ should be understood as a projector on the slow modes within $n=0$ Floquet sector, Eq.~(\ref{EqFloquet}). Therefore $\hat{P} \cl{L}_1 \rho_0=0$ and $\hat{P}\cl{L}_1 \cl{L}_0^{-1}\cl{L}_1\rho_0=\hat{P}\cl{L}_1 \hat{Q}(\hat{Q}\cl{L}_0\hat{Q})^{-1}\hat{Q}\cl{L}_1\rho_0$.

\subsection{Numerical construction of $\rho_0$}\label{numRho0}

In cases where the relevant conservation laws are not known, one can construct for finite systems
also the zeroth order result by brute force using exact diagonalization~\cite{lange17}.

Using the exact eigenstates of $H_0$, $H_0 |n\rangle=E_n^0  |n\rangle$, the set of conservation laws is given by  
\begin{equation}
\mathcal Q=\{ |n\rangle \langle m|\  {\rm with}\  E_n^0=E_m^0\}.
\end{equation}
\comm{To use these conservation laws  in Eq.~\eqref{EqP} which is written for Hermitian conservation laws, one has to construct the corresponding Hermitian operators  $\ket{n}\bra{m}+\ket{m}\bra{n} $ and $(\ket{n}\bra{m}-\ket{m}\bra{n})/i$ for $n \neq m$.
}
Denoting elements of $\mathcal Q$ by $\mathcal Q_i \in \mathcal Q$, one can write $\rho_0=\sum \alpha_i \mathcal Q_i$ which fulfills by constuction the condition $\cl{L}_0 \rho_0=0$. In this case Eq.~\eqref{EqRho0} is a linear equation for the parameters $\alpha_i$ which can be solved by finding the kernel of the matrix 
\begin{equation}
\cl{L}^{\mathcal Q}_{nm}=\tr \,[\mathcal Q_n^\dagger \cl{L}_1 \mathcal Q_m].
\end{equation}
Similarly, the exact solution of Eq.~\eqref{EqUniPert} is obtained by finding the eigenvector with eigenvalue 0 of
\begin{equation}
\cl{L}^{\mathcal Q}_{nm}=-\tr\, [\mathcal Q_n^\dagger \cl{L}_1 \cl{L}_0^{-1} \cl{L}_1\ \mathcal Q_m].
\end{equation} 
Note that the dimension of $\cl{L}^{\mathcal Q}_{nm}$ is much smaller than the dimension of the full Liouvillian super-operators. For a spin-1/2 XXZ Heisenberg chain with $L$ sites in the presence of an external magnetic field, we found in Ref.~\cite{lange17} that the dimension of $\mathcal Q$ is approximately $2\cdot 2^L$ (the factor $2$ arises from degeneracies in the spectrum), to be compared with $(2^L)^2=4^L$, the dimension of density matrices on which the Liouville super-operators act. 

A tricky, unresolved question is under what condition the construction above converges in the thermodynamic limit. For example, there could in principle be situations where a conservation law exists only in the thermodynamics limit but not in the finite size system. A related question is whether the limits $\epsilon \to 0$ and $L \to \infty$ commute or not.
 
When using the set $\mathcal Q$ of conserved quantities, the projector $\hat{P}$ has to be replaced by $\tilde P$,
\begin{equation}\label{EqPtilde}
\tilde{P}X=
\sum_{n,m} \ave{n|X|m}\delta_{E_n^0,E_{m}^0} \ \ket{n}\bra{m} 
\end{equation}
projecting on the density operator subspace spanned by $|n\rangle\langle m|, \ E_n^0=E_{m}^0$. 
This type of projector has been used also in other perturbative studies of Liouville dynamics \cite{benatti11,znidaric15} implementing degenerate perturbation theory for super-operators in case of weak openness.

\section{Perturbation theory}\label{SecPert}
Projectors $\hat{P},\hat{Q}$ are  necessary to develop the perturbation theory of corrections to $\rho_0$, given to all orders by $\delta\rho$ in Eq.~(\ref{EqDeltaRho}).
Below we present a controlled expansion of the inverse $\cl{L}^{-1}$, which removes possible singularities and separates different orders $\mathcal{O}(\epsilon^n)$ within the steady state. 
With $\parallel$ and $\perp$ we denote the density matrix subspaces, which are in the image of $\hat{P}$ and $\hat{Q}$, respectively, $\delta \rho=\hat P \delta \rho +\hat  Q \delta \rho=\delta\rho_\|+\delta\rho_\perp$.
Note that only the tangential component  is relevant for the expectation values of conserved operators $C_i$, 
\begin{align}
&\ave{C_i}
=\tr\big(C_i (\rho_0+\delta\rho)\big)
=\tr\big(C_i (\rho_0+\delta\rho_\parallel)\big)
\end{align}
as $\hat Q^\dagger C_i=0$ for conserved quantities. $\delta\rho_\perp$ does, however, affect expectation values of other operators.

\subsection{Markovian perturbation}\label{SecMarkPert}
First we consider Liouvillians where the perturbations break all relevant conservation laws already to linear order in  $\cl{L}_1 $ such that  $\hat{P}\cl{L}_1 \hat{P}$ has no zero modes. 
\comm{Otherwise the procedure explained in Sec.~\ref{SecPertUni} is needed. 
} 

Our goal is to expand $\delta\rho=- \cl{L}^{-1}\cl{L}_1\rho_0$, Eq. (\ref{EqDeltaRho}), using that we determined $\rho_0$ from the condition $\hat P \cl{L}_1\rho_0=0$, Eq.~\eqref{rho0Pro1}. 
With this definition, we obtain
\begin{align}\label{deltaRhoAll}
\delta\rho=- \cl{L}^{-1} \hat Q \cl{L}_1\rho_0.
\end{align}
\comm{where the Liouville inverse has to be interpreted using regulator as in Eq.~\eqref{EqDeltaRho}, see also App.~\ref{AppRegularize}.}
Due to its conservation laws, $\cl{L}_0^{-1}$, is singular. To avoid these singularities, we 
expand $\cl{L}^{-1} $ around  $\cl{L}_0 + \hat{P}\cl{L}_1\hat{P}$ using
 $\cl{L}_1=(\hat{P}+\hat Q)\cl{L}_1 (\hat{P}+\hat Q)$ and that \comm{$\hat{P}\cl{L}_1\hat{P}$ is invertible in P subspace},
\begin{align}
&\cl{L}^{-1} \hat{Q}=(\cl{L}_0 + \cl{L}_1)^{-1} \hat{Q}
=(\cl{L}_0 + \hat{P} \cl{L}_1 \hat{P})^{-1} \label{EqExpansion}\\
&\times\hspace{-0.09cm}\sum_{n=0}^\infty\hspace{-0.07cm}
\big[\hspace{-0.07cm}-\hspace{-0.07cm}(\hat{P} \cl{L}_1 \hat{Q} + \hat{Q} \cl{L}_1 \hat{P} + \hat{Q} \cl{L}_1 \hat{Q})(\cl{L}_0 + \hat{P} \cl{L}_1 \hat{P})^{-1}\big]^n \hat{Q}.\notag
\end{align}
For power counting in $\epsilon$, it is important to realize that 
\begin{align}\label{slowDyn}
(\cl{L}_0 + \hat{P} \cl{L}_1 \hat{P})^{-1} \hat P& =(\hat{P} \cl{L}_1 \hat{P})^{-1} \hat P\sim \mathcal{O}(1/\epsilon)\\
(\cl{L}_0 + \hat{P} \cl{L}_1 \hat{P})^{-1} \hat Q&=(\hat Q \cl{L}_0 \hat Q)^{-1} \hat Q \sim \mathcal{O}(1)  \nonumber
\end{align}
\comm{where we have used $\hat{P}\cl{L}_0=\cl{L}_0\hat{P}=0$.} This reflects that the dynamics in the subspace of approximately conserved quantities is slow, while it is fast in the perpendicular space. 
Note that $(\hat{P} \cl{L}_1 \hat{P})^{-1}$ and $(\hat Q \cl{L}_0 \hat Q)^{-1}$ in Eq.~\eqref{slowDyn} should be interpreted as {\it inverses within the $P$ and $Q$ subspaces}, respectively, where $\cl{L}_0$ and $\cl{L}_1$ are invertible.
\comm{One first performs projection and then the inversion within the subspace that dynamics was projected on.}
Moreover, Eq.~(\ref{slowDyn}) gives an alternative derivation of Eq.~\eqref{rho0Pro1}: the steady state $\rho_0$ has to fulfill $\hat{P}\cl{L}_1\rho_0=0$, otherwise corrections 
$\delta\rho \sim \cl{L}^{-1} \hat{P} \cl{L}_1\rho_0 \sim 1$ which contradicts our perturbative approach in the limit $\epsilon\to 0$, see App.~\ref{AppUniGGE}.

The combination of Eq.~\eqref{deltaRhoAll}, \eqref{EqExpansion} and \eqref{slowDyn} allows to obtain a straightforward expansion of $\delta \rho$ in powers of $\epsilon$.
The steady state density matrix has a distinct structure in the tangential ($\parallel$) and the perpendicular ($\perp$) subspace in all orders, therefore one has to consider the contributions from the two subspaces separately. 
To linear order in $\epsilon$ we obtain
\begin{align}
\delta \rho &\approx \delta\rho_{1,\parallel} + \delta\rho_{1,\perp}+ \mathcal{O}(\epsilon^2), \notag \\
\delta\rho_{1,\parallel}&=
(\hat{P}\cl{L}_1 \hat{P})^{-1} \ \hat{P}\cl{L}_1 \hat{Q} \ \cl{L}_0^{-1} \ \hat{Q}\cl{L}_1 \rho_0,\label{EqLinRespP}\\
\delta\rho_{1,\perp}&=-\cl{L}_0^{-1} \ \hat{Q} \cl{L}_1 \rho_0. \label{EqLinRespQ} 
\end{align}
For brevity we use notation $\hat{Q} (\hat{Q}\cl{L}_0 \hat{Q})^{-1} \hat{Q}=\hat{Q} \cl{L}_0^{-1} \hat{Q}$, where $\cl{L}_0^{-1}$ should be understood as an inverse within $Q$ subspace only. 
We would like to stress again, that  $\rho_0$ already contains effects of order $\epsilon^0$.
But after $\rho_0$ has been correctly chosen, $\delta \rho$ does indeed vanishes for $\epsilon \to 0$,  $\delta \rho \sim \epsilon$. In the $\|$ space, one has, however, to expand to {\em second} order in $\cl{L}_1$ to
obtain the correction of order $\epsilon$ due to the presence of the  $(\hat{P}\cl{L}_1 \hat{P})^{-1}$ term. This contributes as $1/\epsilon$ factor, resulting in a term proportional to $\epsilon^2/\epsilon=\epsilon$. In the perpendicular space no such issue arises. The fact that perturbation theory for steady states has a different structure compared to perturbation theory in thermal equilibrium is well known, see, e.g. Ref.~\cite{rosch03}, and can simply be understood from the fact that rate equations determining the steady state have to be readjusted in the presence of perturbations.

Note that the change $\delta \rho_\|$ can in principle be absorbed in a redefinition of the Lagrange parameters, while $\delta \rho_\perp$ describes contributions which cannot be captured by a GGE ensemble.

For concrete calculations one needs to compute
\begin{align}\label{EqPLPinv}
(\hat{P}\cl{L}_1\hat{P})^{-1}\hat P X
&=-\sum_{pr}\frac{\partial\rho_0}{\partial\lambda_p} (M^{-1})_{pr} \ \tr[C_r X],
\end{align}
where $M^{-1}$ is the inverse of matrix $M$ with components 
$M_{pr}=-\tr[C_p \, \cl{L}_1 (\partial\rho_0/\partial\lambda_r)]$, see App.~\ref{AppProj}.
Then, for example, the change of the expectation value of a conserved operator $C_i$ is given linearly in $\epsilon$ by, $\delta \ave{C_i}\approx \ave{C_i}_1$
\begin{align}
\ave{C_i}_1
&=\tr\left[ C_i (\hat{P}\cl{L}_1\hat{P})^{-1} \hat{P}\cl{L}_1 \hat{Q} \cl{L}_0^{-1} \hat{Q} \cl{L}_1\rho_0 \right] \label{EqLinRespCP}\\
&=\sum_{jk} \chi_{ij} \, (M^{-1})_{jk} 
\, \tr\big[C_k\cl{L}_1 \hat{Q}\cl{L}_0^{-1} \hat{Q}\cl{L}_1\rho_0\big]\label{EqLinRespC}
\end{align}
\begin{figure}[!b]
\center \includegraphics[width=\linewidth]{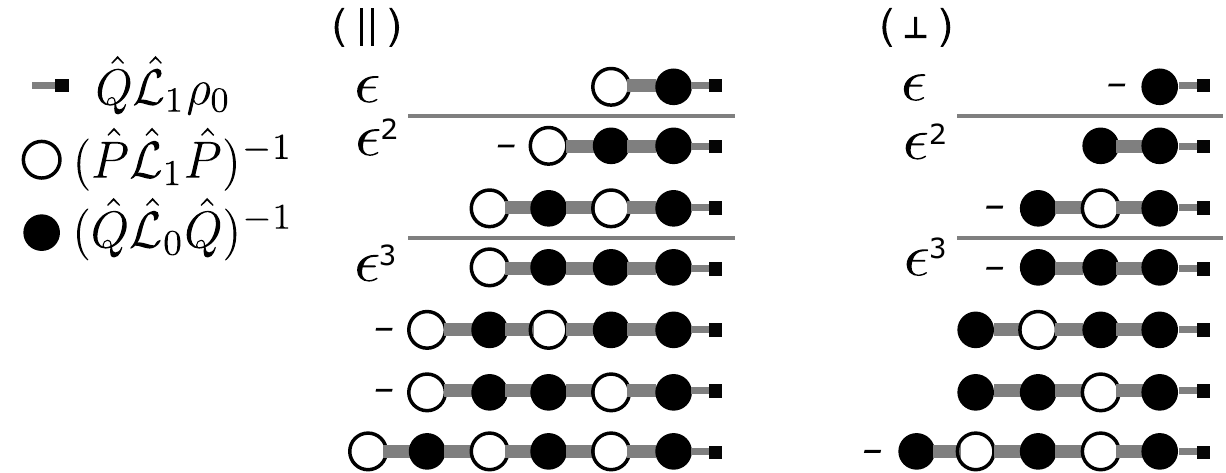}
\caption{Diagrammatic depiction of the structure of corrections to the zeroth order density matrix $\rho_0$. One draws all possible combinations of open and filled circles starting to the right with a filled circle connected to a small square representing $\rho_0$. Then one eliminates all diagrams with a direct connection of two open circles. The order of the diagram is given by the number of filled circles, the sign by the total number of lines. The number of terms to order $\epsilon^n$ is given by $2^{n}$.
Note that the corrections to order $n$ in the perpendicular and parallel sector are simply related by the relation $\delta \rho_{n,\|} =-(\hat P \cl{L}_1 \hat P)^{-1} (\hat P \cl{L}_1 \hat Q) \, \delta \rho_{n,\perp}$.  
\label{fig0}}
\end{figure}
Higher order corrections can be obtained in a straightforward way from the Taylor expansion, Eq.~(\ref{EqExpansion}), in combination with the rules of Eq.~\eqref{slowDyn} which show how the inverse $(\cl{L}_0 + \hat{P} \cl{L}_1 \hat{P})^{-1}$ has to be evaluated.
\comm{ Eq.~\eqref{EqLinRespCP} and its higher order generalization can be rewritten in terms of standard correlation functions as we discuss in App.~\ref{AppCorr} where concrete formulas for $\ave{C_i}_1$ and $\ave{C_i}_2 \sim \mathcal{O}(\epsilon^2)$ are given. 
}

 In Fig.~\ref{fig0} we give a graphical representation of the relevant terms in the tangential ($\parallel$) and the orthogonal ($\perp$) subspace. As written on the left of the figure, the two types of inverse Liouville super-operators in $P$ and $Q$ space are shown as white and black circles, respectively. There are three types of links, $\hat Q \cl{L}_1 \hat Q$, $\hat P \cl{L}_1 \hat Q$ and $\hat Q \cl{L}_1 \hat P$, connecting black and black, white and black, and black and white circles, respectively. This simply follows from the rule that  $\hat P \hat Q=0=\hat Q \hat P$.

Starting from $\hat Q \cl{L}_1 \rho_0$ (small square with thin line) to the right, one attaches circles and lines in all allowed combinations. The color of the last circle on the left, determines whether the density matrix operator is in $P$ or $Q$ space.
The power in $\epsilon$ is determined by the number of lines minus the number of white circles as
$(\hat{P}\cl{L}_1\hat{P})^{-1} \sim 1/\epsilon$. The sign is simply given by $(-1)^{N_L}$ where $N_L$ is the number of lines.

\comm{We have checked that the formulas discussed above can alternatively be derived starting
from projection-operator based time-dependent perturbation theories in the version discussed by Breuer and Petruccione \cite{breuer02,Koide02}.}

As a remark we should point out that there is no need for any additional normalization since $\tr\,\delta\rho=0$ which is guaranteed by the regularization of the inverse operators, see Appendix \ref{AppRegularize}. 


\subsection{Missing conservation laws}

Above, we have assumed that all relevant conservation laws of the unperturbed system are known and have been included in the construction of the GGE (or that $\rho_0$ has been constructed numerically, see Sec.~\ref{numRho0}). It may, however, happen that either not all conservation laws are known or that it is just technically impossible to include them all. This might, for example, be the case in integrable systems where an infinite number of conservation laws exists.

In this case one may try to approximate system by a {\em truncated} GGE, 
\begin{eqnarray}
\rho^{[t]}_0=\frac{\exp\left[-\sum_i \lambda_i^{[t]} C_i^{[t]}\right]}{\tr [ e^{-\sum_i \lambda_i^{[t]} C_i^{[t]}}]}
\end{eqnarray}
considering only a finite subset of conserved operators. We denote by $\hat P_t$ the projector on the space  tangential to the truncated GGE, and by $\chi^{[tt]}$ the matrix of susceptibilities, all defined as above.

 A selection criterion for the truncated space could be, for example, to consider only the most local conservation laws, containing less than a certain number of derivatives in a continuum model or having a support of less than a certain number of sites for a lattice model. A truncated GGE was, for example, used in Ref.~\cite{lange17}. In the following we will consider two questions: (i) How can one compute perturbatively the effects of the 'other' conservation laws, and (ii) how does the perturbation theory developed above signal the presence of missing conservation laws.

In the following we denote the missing conservation laws not included in the truncated GGE by $C_i^{[m]}$. Similarly, we define susceptibility matrices $\chi^{[mm]}_{ij}=\langle C_i^{[m]}C_j^{[m]}\rangle_{_{0,\textrm c}}$ and $\chi^{[mt]}_{ij}=\langle C_i^{[m]}C_j^{[t]}\rangle_{_{0,\textrm c}}$, where $\ave{\cdot}_{_0}$ stands for expectation values with respect to $\rho_0^{[t]}$.
Without loss of generality, we assume that they are orthogonal to the $C_i^{[t]}$,
$\chi^{[mt]}_{ij}=0$ for all $i,j$ (if this is not the case, one can simply replace them by ${\hat Q_t}^\dagger C_i^{[m]}$ where $\hat{Q}_t=\hat 1-\hat P_t$). We define the projector on the tangential space corresponding to the missing conservation laws by
\begin{align}
 \hat{P}_m \, X
=
 &-\sum_{i,j} \frac{\partial \rho_0}{\partial \lambda_i^{[m]}}\Big|_{\rho_0^{[t]}} \ (( {\chi^{[mm]}})^{-1})_{ij} \ \tr[C_j^{[m]} X]
\end{align}
with $(\partial \rho_0/\partial \lambda_i^{[m]})\big|_{\rho_0^{[t]}}=-(C^{[m]}_i-\langle C^{[m]}_i \rangle_{_0}) \rho_0$. The sum $\hat P= \hat{P}_t+\hat{P}_m$ is then the projector on the tangential space spanned by all conservation laws.

Assuming that the missing conservation laws give only a small correction to the Lagrange parameters $\lambda_i$, we can Taylor expand in $\delta\lambda_k=\lambda_k-\lambda_k^{[t]}$ 
using
\begin{align}
\rho_0&=\rho_0^{[t]} + \delta\rho_0, \\
\delta\rho_0&=-\rho_0^{ [t]}\sum_k \delta\lambda_k \bar{C}_k, \ 
\bar{C}_k=C_k-\ave{C_k}_{_0} \nonumber
\end{align}
where the $k$ sum includes both the $C_i^{[t]}$ and $C_i^{[m]}$.
From Eq.~\eqref{EqRho0}, we obtain directly a matrix equation $A \boldsymbol{\delta\lambda}= \boldsymbol{a}$  for $\delta \lambda_i$ solved by $\boldsymbol{\delta\lambda}=A^{-1} \boldsymbol{a}$  which is written in components as
\begin{align}
\left( \begin{array}{c}
\boldsymbol{\delta\lambda}^{[t]}\\
\boldsymbol{\delta\lambda}^{[m]}
\end{array} \right)
&=\left( \begin{array}{cc}
A^{[tt]} & A^{[tm]} \\
A^{[mt]} & A^{[mm]}
\end{array} \right)^{-1}
\left( \begin{array}{c}
0 \\
\boldsymbol{a}^{[m]} 
\end{array} \right)
 \label{EqDeltaLambda}\\ 
(\boldsymbol{a}^{[m]})_i &=  \ \tr[C_i^{[m]} \cl{L}_1 \rho_0^{[t]}]
 = \langle \dot C_i^{[m]}\rangle_{_0}
\nonumber \\
A^{[I J]}_{ij}
&= \tr[C^{[I]}_i \cl{L}_1 (\bar{C}^{[J]}_j\rho_0^{[t]})]
=\ave{\dot C^{[I]}_i \bar{C}^{[J]}_j} _{_{0,\textrm c}} \nonumber
\end{align}
Note that $A^{[mm]}$ is equivalent to matrix $M$ in Eq.~\eqref{EqPLPinv}.

From the change of the Lagrange parameters, one can also directly calculate the change of observables. We are mainly interested in the change of $\ave{\delta C_i^{[t]}}_{_0}$
\begin{align}\label{deltaCmissing}
\ave{\delta C_i^{[t]}}_{_0}
&=-\left(\chi^{[tt]} \big(A^{-1}\big)^{[tm]}  \ \boldsymbol{a}^{[m]}\right)_i \\
&\approx \left(\chi^{[tt]} \big(A^{[tt]}\big)^{-1} A^{[tm]} \big(A^{[mm]}\big)^{-1} \ \boldsymbol{a}^{[m]}\right)_i \nonumber
\end{align}
In the last line we did an extra approximation, assuming that the relevant matrix elements between
included and missing conservation laws, $C^{[t]}_i$ and  $C^{[m]}_j$, are small, allowing to expand $A^{-1}$ in $A^{[tm]}$. We conclude that the effect of the missing conservation laws is small if either $ \langle \dot C_i^{[m]}\rangle_{_0}$, the changes of missing conservation laws, are small and/or if the dynamical coupling  $A^{[tm]} _{ij}=\ave{\dot C^{[t]}_i \bar{C}^{[m]}_j} _{_{0,\textrm c}}$ is small. The latter susceptibility describes how a change of the Lagrange parameter $\lambda^{[m]}_j$ induces a finite $\ave{\dot C^{[t]}_i}_{_0}$.

Fully equivalent to Eq.~\eqref{deltaCmissing}, one can write everything using projection super-operators $\hat{P}_t, \hat{P}_m$
\begin{align}\label{deltaC2}
&\ave{\delta C_i^{[t]}}_{_0}
=-\tr \left[C_i^{[t]} \hat{P}_t \, (\hat{P} \cl{L}_1 \hat{P})^{-1} \hat{P}_m \cl{L}_1 \rho_0^{[t]}\right] \\
&\approx \tr \left[C_i^{[t]} (\hat{P}_t \cl{L}_1 \hat{P}_t)^{-1} (\hat{P}_t \cl{L}_1 \hat{P}_m) (\hat{P}_m \cl{L}_1 \hat{P}_m)^{-1} \hat{P}_m \cl{L}_1 \rho_0^{[t]}\right]\notag
\end{align}

Above, we have calculated effects which occur in cases when the problem was not solved accurately to order $\epsilon^0$ because not all conservation laws of $H_0$ had been considered.
If a theory is not treated correctly to order $\epsilon^0$, any perturbative treatment in $\epsilon$ should signal this by the presence of divergencies. In the following, we will show that this is the case within our approach and we will identify directly the origin of these divergencies.

Technically, the divergence in the perturbative formula \eqref{EqLinRespCP} arises from the inverse super-operator $\hat Q_t (\hat Q_t\cl{L}_0 \hat Q_t)^{-1} \hat Q_t$ when only a truncated set of conservation laws is used. As $\hat P_m$ is the projection operator on the missing conservation laws, we can calculate the  divergent contribution from
\begin{equation}
\hat Q_t (\hat Q_t\cl{L}_0\hat Q_t)^{-1} \hat Q_t=-\frac{\hat P_m}{\eta}+O(\eta^0)\label{divL00}
\end{equation}
where $\eta$ is the regulator used to define $\cl{L}_0^{-1}$, see appendix~\ref{AppRegularize}.
We note that exactly the same divergences lead to the occurrence of infinities in the current response
of, e.g.,  integrable systems. As already discussed in Sec.~\ref{SecForces}, the latter are described by Drude weights \cite{mazur69,suzuki71,zotos97} obtained from $\langle (\hat P^\dagger J) J  \rangle_{_{c}}$.
 
The divergent contribution  to $\ave{\delta C^{[t]}_i}$ within $O(\epsilon)$ perturbation theory around the truncated GGE is obtained from Eqs.~\eqref{EqLinRespCP}  and \eqref{divL00} as
\begin{align}
\ave{&\delta C^{[t]}_i}\approx& \label{divPT}\\
&-\tr\left[ C^{[t]}_i (\hat{P}_t\cl{L}_1\hat{P}_t)^{-1} (\hat{P}_t\cl{L}_1 \hat{P}_m) \frac{1}{\eta} \hat{P}_m \cl{L}_1\rho_0 \right]+O(\eta^0) \nonumber
\end{align}
This contribution is of order $\epsilon/\eta$. Comparing Eq.~\eqref{divPT} and the second line of Eq.~\eqref{deltaC2} one observes that the two terms are identical if one replaces $(-\eta)$ by $\hat{P}_m \cl{L}_1 \hat{P}_m$ which is linear in $\epsilon$. Divergencies in perturbation theory can therefore be used to detect missing conservation laws. If, however, the weight of the divergent contribution is small, one can expect that adding the missing conservation laws will have little effect.

\subsection{Unitary driving, 
$\hat{P}\cl{L}_1\hat{P}= 0$} \label{SecPertUni}

In the discussion given above, we assumed that $\hat P \cl{L}_1 \hat P$ is finite and invertible within $P$ subspace.
For an important class of perturbations, arising from a (Floquet) Hamiltonian $H_1$, this is not the case as  $\tr(C_i [H_1,\partial \rho_0/\partial \lambda_j])=0$ due to the cyclic property of the trace
and one therefore finds $\hat{P}\cl{L}_1\hat{P}= 0$.
In this case, the rates with which the $C_i$ change are second order in $\epsilon$, as is well-known from Fermi's golden rule.

For $\hat{P}\cl{L}_1\hat{P}= 0$ the exact inverse of the Liouvillian in the $\hat P$ sector is given by $\hat P \cl{L}^{-1} \hat P=-\hat P (\hat{P} \cl{L}_1 \hat Q (\hat Q \cl{L} \hat Q)^{-1} \hat Q \cl{L}_1 \hat{P})^{-1}\hat P $, see appendix \ref{AppProjInv}. In the limit of small $\epsilon$, we therefore find 
 \begin{eqnarray}\label{EqPLiP}
 \hat P \cl{L}^{-1} \hat P= \hat P(\hat P \cl{L}_2 \hat P)^{-1}\hat P + \mathcal O (\epsilon^{-1})
 \end{eqnarray}
 where 
 \begin{eqnarray}
 \cl{L}_2=
 -\cl{L}_1 \hat Q \, (\hat Q \cl{L}_0 \hat Q)^{-1} \, \hat Q \cl{L}_1.  \label{L2def}
 \end{eqnarray}
$\hat{P}\cl{L}_2 \hat{P}\propto \epsilon^2$ is an effective Lindblad super-operator acting in the $ P$ space.
 With this notation the condition for $\rho_0$,  Eq.~\eqref{EqUniPert}, takes the form $\overline{\tr[C_i \cl{L}_2 \rho_0]}=0$, or, equivalently, Eq.~\eqref{rho0Pro2}, is written as
 \begin{eqnarray}\label{rhoCond3}
 \hat P \cl{L}_2 \rho_0 = 0.
 \end{eqnarray}
 where one should keep in mind that $\hat{P}$ project onto non-oscillatory density matrices only.
 Perturbation theory can now be derived in a straightforward way by a Taylor expansion of
  \begin{eqnarray}\label{splitting}
(\cl{L}_0+\cl{L}_1)^{-1} =\big((\cl{L}_0+\hat{P}\cl{L}_2 \hat{P}) +( \cl{L}_1-\hat{P}\cl{L}_2 \hat{P})\big)^{-1} 
 \end{eqnarray}
in the second term, with the inverse applied as in equation (\ref{EqDeltaRho})  for $\delta \rho$.
 
\begin{figure}[!b]
\center \includegraphics[width=\linewidth]{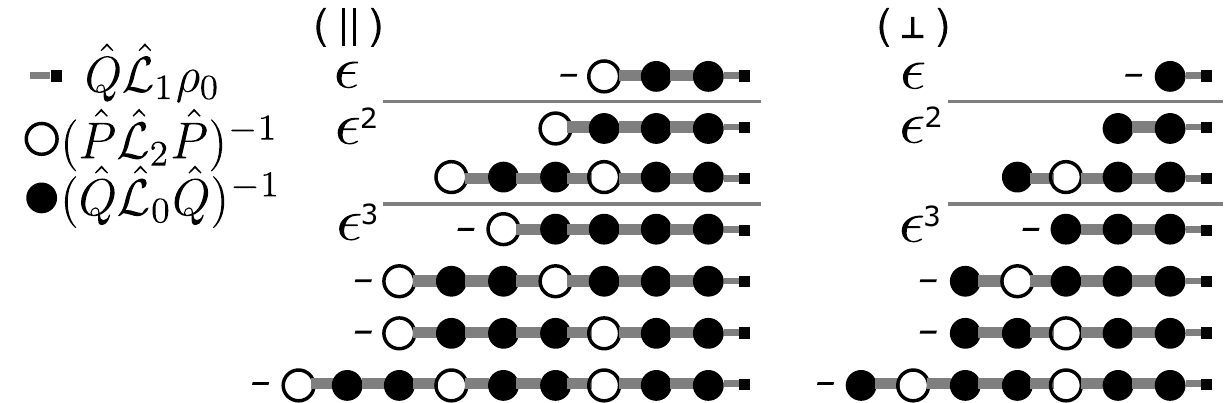}
\caption{Diagrammatic depiction of the corrections $\delta \rho$ to the zeroth order density matrix $\rho_0$ for cases where $\hat P \cl{L}_1 \hat P=0$ (unitary driving). One first draws all possible combinations of open and filled circles starting to the right with a filled circle connected to $\rho_0$ (small square). Then one eliminates all diagrams with neighboring open circles,  all with the combination open-filled-open, and finally also the combination open-filled-$\rho_0$ (small square). The order is given by the number of filled minus the number of open circles, the sign by the total number of circles. The number of terms to order $\epsilon^n$ is $2^{n }$. Some diagrams do, however, vanish for monochromatic perturbations, see appendix~\ref{monochromatic}.
\label{figDiagramsNew}}
\end{figure}

 One can, again, derive diagrammatic rules to calculate corrections to order $\epsilon^n$, see Fig.~\ref{figDiagramsNew}. Compared to the previous case, shown in Fig.~\ref{fig0}, there are only two changes. First, $(\hat P \cl{L}_1 \hat  P)^{-1}\sim \mathcal O(\epsilon^{-1})$ (open circle) is replaced by $(\hat  P \cl{L}_2 \hat  P)^{-1}\sim \mathcal O(\epsilon^{-2})$ and, second, two sets of diagrams do not contribute any more as they either cancel due to the $-\hat{P}\cl{L}_2\hat{P}$ in Eq.~\eqref{splitting} or are set to zero by the condition \eqref{rhoCond3} for $\rho_0$. Explicitly, there are no diagrams with neighboring open circles, with the combination open-filled-open, and finally also the combination open-fillled-$\rho_0$ (small square). The cancellation of these diagrams is also a necessary condition for the series expansion in $\epsilon$ to be valid.
 In Fig.~\ref{figDiagramsNew}, we show all remaining diagrams up to  $\mathcal O(\epsilon^3)$ and describe in the figure caption the corresponding diagramatic rules.

 We would like to finish the section by pointing out that there can be more complex situations \cite{jung06,lenarcic15} where perturbation theory to order $\epsilon^2$ does not fix $\rho_0$ and where accordingly $\hat{P}\cl{L}_2\hat{P}$ vanishes, see discussion at the end Sec.~\ref{SecCondGGE}.  Or even more general, there can be cases where Hamiltonian and Lindblad perturbations occur on equal footing or where some approximate conservation laws change by processes to order $\epsilon$ and others by order $\epsilon^2$.  We believe that in all these cases, one can generalize the approach described above: one defines projectors on subspaces governed by similar time-scales (same power of $\epsilon$), identifies the leading-order dynamics ($\hat{P}\cl{L}_2\hat{P}$ in the case discussed above)  in each subsector perturbatively, uses this to fix $\rho_0$ in analogy to Eq.~(\ref{rhoCond3}), and performs then a Taylor expansion  using the analog of Eq.~(\ref{splitting}) as a starting point.

\section{Example: Interacting Fermions with particle-gain and loss}\label{example}
\comm{
In the following we will give two examples where the perturbation theory developed above applies. We consider interacting fermions 
driven out of equilibrium by a weak coupling to the environment. As perturbation we choose processes which lead to a particle loss with rate $\epsilon l_n$ and a gain of particles with rate $\epsilon g_n$. In quantum optics experiments, the quasiparticles could be excitations of atoms (or cavities) and gain and loss is realized by emission and absorption of light. The same situation arises in experiments on exciton condensation, magnon condensation or in photon BECs \cite{butov01,kasprzak06,balili07,klaers10,klaers12,schmitt14,demokritov06,chumak09}.
Also in experiments with ultracold atoms, loss processes arise when atoms absorb photons from external laser beams, kicking them out of their trap. However, most cold-atom experiments do not include processes where the lost atoms are replenished and therefore a true steady state, the main focus of our study, cannot be reached.

We first study in Sec.~\ref{SecExact} a small finite-size system where we can compare directly perturbation theory and exact solution. Here it is important to realize that the applicability of perturbation theory for small systems (in the limit where perturbations are small compared to the level spacing) does not guarantee the validity of perturbation theory for systems close to the thermodynamic limit (perturbations large compared to the level spacing but small compared to internal equilibration rates). We therefore  study in Sec. \ref{SecBoltz} the thermodynamic limit by considering a regime where the Boltzmann equation can be applied.
A numerical investigation of the thermodynamic limit in a quantum approach is beyond the scope of the present paper and is left for future studies \cite{inpreparation}.

\subsection{Lindblad dynamics in a small system}\label{SecExact}

We consider a model where the Hamiltonian dynamics of the unperturbed system is given by
\begin{equation}
	H_0=\sum_{n=1}^L e_n c^\dagger_n c_n + U \sum_{n_1>n_2,n_3<n_4} c^\dagger_{n_1}c^\dagger_{n_2}c_{n_3}c_{n_4}
\end{equation}
with $e_n=n/L$.
Gain and loss processes are described by the Lindblad operators 
$L^g_n=c^\dagger_n$ and $L^l_n=c_n$ such that 
\begin{eqnarray}
	\cl{L}_1 &=&\epsilon (\hat D_g +D_l )\\
	\hat D_g &=&\sum g_n \left(L^g_n \rho {L^g_n}^\dagger-\frac{1}{2} \{{L^g_n}^\dagger{L^g_n},\rho\} \right) \nonumber \\
	\hat D_l &=&\sum l_n \left(L^l_n \rho {L^l_n}^\dagger-\frac{1}{2} \{{L^l_n}^\dagger{L^l_n},\rho\} \right) \nonumber
\end{eqnarray}
with 
\begin{align}\label{EqGainLossRate}
& g_n=\frac{1}{4}, \quad l_n = e_n
\end{align}

\begin{figure}[b!]
\includegraphics[width=.83\linewidth]{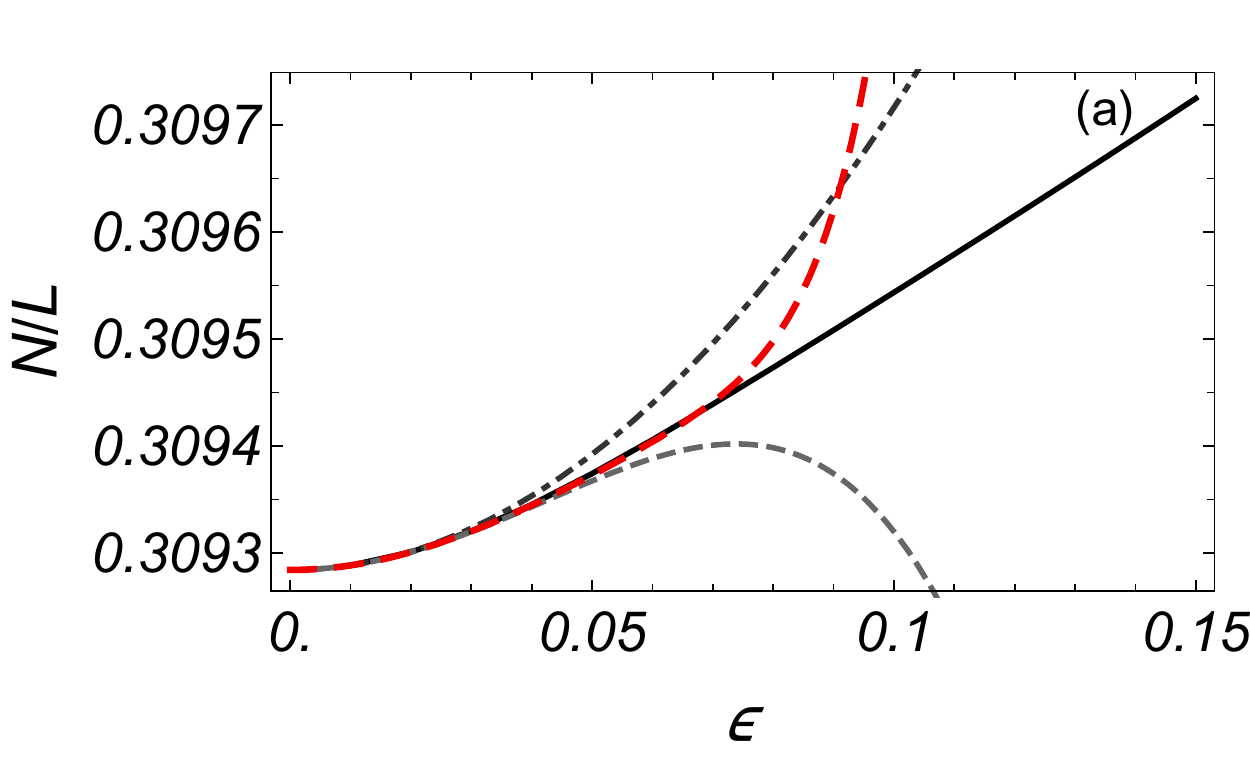}\\
\vspace{-0.2cm}
\includegraphics[width=.83\linewidth]{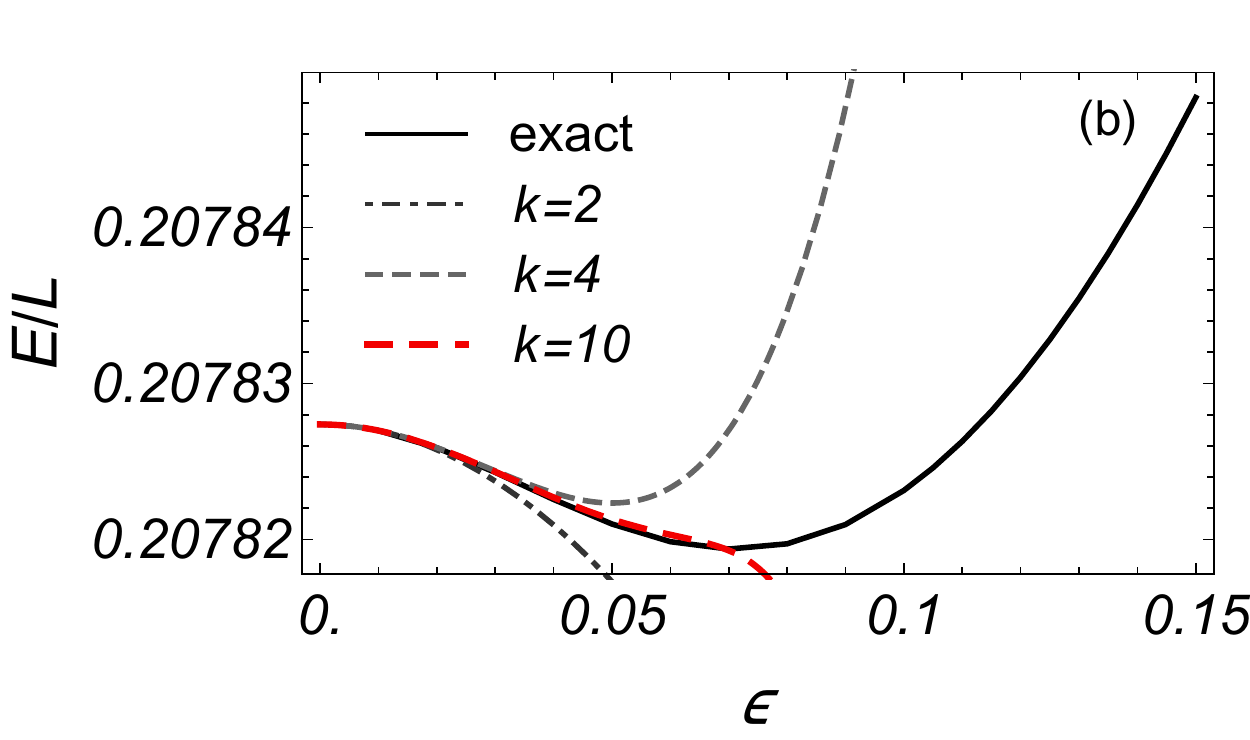}
\caption{(Color online) Expectation values of (a) particle and (b) energy density as a function of perturbation strength $\epsilon$, calculated from the exact steady state density matrix (solid line) or using our perturbation theory up to $k$-th order in $\epsilon$ on systems size $L=4$, and interaction strength $U=0.3$. \label{FigSmallExact}}
\end{figure}

The loss rate $\epsilon l_n$ depends on energy. This mimics a situation where high-energy states have a higher probability to evaporate than low-energy states, thus implementing a cooling mechanism. Note that  $\epsilon$, the common prefactor of $\hat D_i$, controls the overall strength of both the heating and cooling terms.

The conservation laws of the small system are just projectors on the eigenstates of $H_0$, $|n\rangle\langle n|$ with $H_0\ket{n}=E_n^0 \ket{n}$. Using the formulas of Secs.~\ref{SecLeading},\ref{SecPert} we can easily solve for the 0th order steady state and determine the perturbations around it. In Fig.~\ref{FigSmallExact} we show the expectation values of particle and energy density as a function of perturbation strength $\epsilon$, calculated from the exact steady state density matrix (solid line) or using our perturbation theory up to $k$-th order in $\epsilon$ on systems size $L=4$, $k=2,4,10$, and interaction strength $U=0.3$. 

The figures show that for this small system with finite level spacing the perturbation theory works for small $\epsilon$. For the chosen model we find both from the exact result and from the perturbation theory that the variations of particle number and energy are tiny. The perturbation theory breaks down for rather small $\epsilon$ of the order of the level spacing.
This is a phenomenon well known from standard perturbation theory where perturbative corrections are inversely proportional to the level spacing. This is, however, dfferent when correlations functions are evaluated in the thermodynamics limit by taking first the limit $L \to \infty$ and then the limit $\eta \to 0$, where $\eta$ is the regulator defined in Eq. (\ref{EqDeltaRho}), see also App. \ref{AppCorr}. While the perturbative correction linear in $\epsilon$ vanishes for the finite system ($\eta$ smaller than level spacing), it is finite in the thermodynamics limit ($\eta$ larger than level spacing).
}

\subsection{Boltzmann dynamics}\label{SecBoltz}
\comm{Now we consider a similar example in the thermodynamic limit where the unperturbed system $H_0$ has only two local conservation laws, energy and particle number.} 
Interactions lead to an equilibration of all other conservation laws. We assume that the system can be described by weakly interacting quasiparticles with energies $e_n$, such that the collision term of a Boltzmann equation captures their equilibration dynamics. As above we consider fermions with a constant density of states, which we discretize using $L$ single-particle states with energies $e_n$ equally spaced between $0$ and $1$ with $e_n=n/L$, $n=1,\dots,L$. The Boltzmann equation takes the form,
\begin{align}
\frac{d f_e}{dt}
&=M [f]_e + \epsilon\,  D[f]_e \label{EqBoltzmann}\\
M [f]_e&=\int_0^{1} de_1 de_2 de_3  \,  \delta(e+e_1-e_2-e_{3})  \notag \\
& \qquad \qquad\qquad \times (\bar{f}_e \bar{f}_{e_1} f_{e_2} f_{e_3}-f_{e} f_{e_1} \bar{f}_{e_2} \bar{f}_{e_3})\notag \\
&= \frac{1}{L^2}
\sum_{i,j,l} (\bar{f}_e \bar{f}_{e_i} f_{e_j} f_{e_l}-f_e f_{e_i} \bar{f}_{e_j} \bar{f}_{e_l}) \ \delta_{e+e_i,e_j+e_{l}} \notag \\
D[f]_e&=-l_e f_e + g_e \bar{f}_e \label{EqBolzLoss}
\end{align}
where $f_{e_n}$ is the occupation function as function of the energy $e_n$ and $\bar{f}_e=(1-f_e)$.
The Kronecker-$\delta$ guarantees energy conservation in each collision process of the discretized model. Note that we consider a model where momentum conservation does not play a role.
For simplicity, we also set all transition rates due to collisions to unity.
\comm{
We keep the same type of perturbations as in the previous example, i.e. particle gain and loss. Instead of Lindblad operators these are now encoded directly in Boltzmann equation through $D[f]_e$, Eq.~\eqref{EqBolzLoss}. Rates for particle gain and loss are the same as in Eq.~\eqref{EqGainLossRate}, i.e., $g_e=1/4,l_e = e$. 
Note that  $\epsilon$, the prefactor of $D[f]_e$, controls the overall strength of both the heating and cooling terms. Numerical calculations are performed for $L=41$.
}

\begin{table}[b!]
\caption {Comparison of semi-classical open Boltzmann dynamics for level occupation function and quantum Liovillian formulation for density
matrix} \label{table} 
\begin{ruledtabular}
\bgroup
\def\arraystretch{1.5}
\begin{tabular}{ l  l }
 Boltzmann & Liouvillian \\
 \hline
 occupation function $f_e$ & density matrix $\rho$\\
$\frac{d f_e}{dt}=M [f]_e + \epsilon\,  D[f]_e$  & $\frac{d \rho}{dt}=\cl{L}_0 \rho + \cl{L}_1 \rho$ \\
 $f_e(t\to \infty)= f_e^0 + \delta f_{e}$ & $\rho(t \to \infty)=\rho_0 + \delta\rho$ \\
 Fermi function: $f_e^0=\frac{1}{1+e^{\beta(e - \mu)}}$ & GGE: $\rho_0 = \frac{e^{-\lambda_i C_i}}{\tr[e^{-\lambda_i C_i}]}$\\
\hline
{\it conservation laws and scalar product} \hspace*{-1cm}& \vspace{-0.1cm}\\
$c_i(e)$ and $q_i^{\beta,\mu}$  & $C_i$ and $\partial{\rho_0}/\partial{\lambda_i}$\\
$\int de   \,  c_i(e) f_e$ & $\tr[C_i\rho]=\ave{C_i}$\\
$\chi_{ij}(\beta,\mu)=-\int de \, c_i(e) \,q_j ^{\beta,\mu}(e)$ & $\chi_{ij}=-\tr[C_i\partial{\rho_0}/\partial{\lambda_j}]$\\
\hline
{\it zeroth order perturbation theory}& \vspace{-0.1cm}\\
$\int de   \,  c_i(e)  \epsilon D[f^0]_e=0$ & $\tr[C_i \cl{L}_1\rho_0]=0$\\
$\hat{P}\big[D[f^0]\big]=0$ & $\hat P (\cl{L}_1\rho_0)=0$\\
\hline
{\it first order corrections}& \vspace{-0.1cm}\\
$\boldsymbol{f}^1_{\perp}=-(QM^{(0)}Q)^{-1} Q \boldsymbol{D}[\boldsymbol{f}^0]$ & $\delta\rho_{1,\perp}=-(\hat{Q}\cl{L}_0\hat{Q})^{-1} \ \hat{Q} \cl{L}_1 \rho_0$\\
$\boldsymbol{f}^1_{\parallel}=(PD^{(1)}P)^{-1}PD^{(1)} Q \, \times$ & $\delta\rho_{1,\parallel}=(\hat{P}\cl{L}_1 \hat{P})^{-1} \hat{P}\cl{L}_1 \hat{Q} \ \times$ \\
$\hspace{1.1cm}\times(QM^{(0)}Q)^{-1} Q\boldsymbol{D}[\boldsymbol{f}^0]$ & $\hspace{2cm}\times  \cl{L}_0^{-1}\hat{Q}\cl{L}_1 \rho_0$
\end{tabular}
\egroup
\end{ruledtabular}
\end{table}

\comm{The perturbation theory derived in the main text can be with some straightforward modifications also applied to the open Boltzmann dynamics. Before we proceed we collect in Table \ref{table} all analogies with the formulation for the Liouvillian dynamics, developed in the main part of the article. Note that some objects are defined later within this section.}

The collision integral $M$ preserves the total energy $E=\sum_{n} e_{n} f_{e_n}$ and the particle number $N=\sum_n  f_{e_n}$ and these are the only conserved quantities in the absence of loss/gain term.
In the following we expand the level occupations $f_e$ in orders of $\epsilon$,
\begin{equation}\label{EqExpand}
f_e(t\to\infty)=\sum_{m} \epsilon^m f^m_e.      
\end{equation}
In correspondence with $\cl{L}_0\rho_{0}=0$ now any Fermi-Dirac distribution $f_e^0$ satisfies $M [f^0]=0$,
\begin{align}
f_e^0(\beta,\mu)=\frac{1}{1+e^{\beta(e - \mu)}} \quad &\leftrightarrow \quad  \rho_0,\quad\notag\\
M [f^0(\beta,\mu)]_e=0  \quad  &\leftrightarrow  \quad  \cl{L}_0\rho_0=0    
\end{align}
and only the perturbation $D$  fixes the parameters $\beta$ and $\mu$, where $\beta$ is the Lagrange parameter of the energy and $-\beta \mu$ is the Lagrange parameter of the particle number.

{\it Zeroth order---} The parameters $\beta$ and $\mu$ of  the steady-state in the $\epsilon\to 0$ limit,   $f^0_e(\beta,\mu)$, are determined by the stationarity of conserved quantities (total energy and particle number) from a set of coupled equations corresponding to Eq.~(\ref{EqRho0}), 
\begin{align}\label{EqBolz0}
\frac{1}{L}\frac{dE}{dt}
&\approx \epsilon  \int de \ e \big(-l_{e} f_{e}^0 + g_{e} (1-f_{e}^0)\big)\stackrel{!}{=}0 \notag\\
\frac{1}{L}\frac{dN}{dt}
&\approx \epsilon \int de \ \big(-l_{e} f_{e}^0 + g_{e} (1-f_{e}^0)\big)\stackrel{!}{=}0
\end{align}
For our  model we find $\beta=2.328$ and $\mu=0.288$.

\begin{figure}[b!]
\includegraphics[width=.83\linewidth]{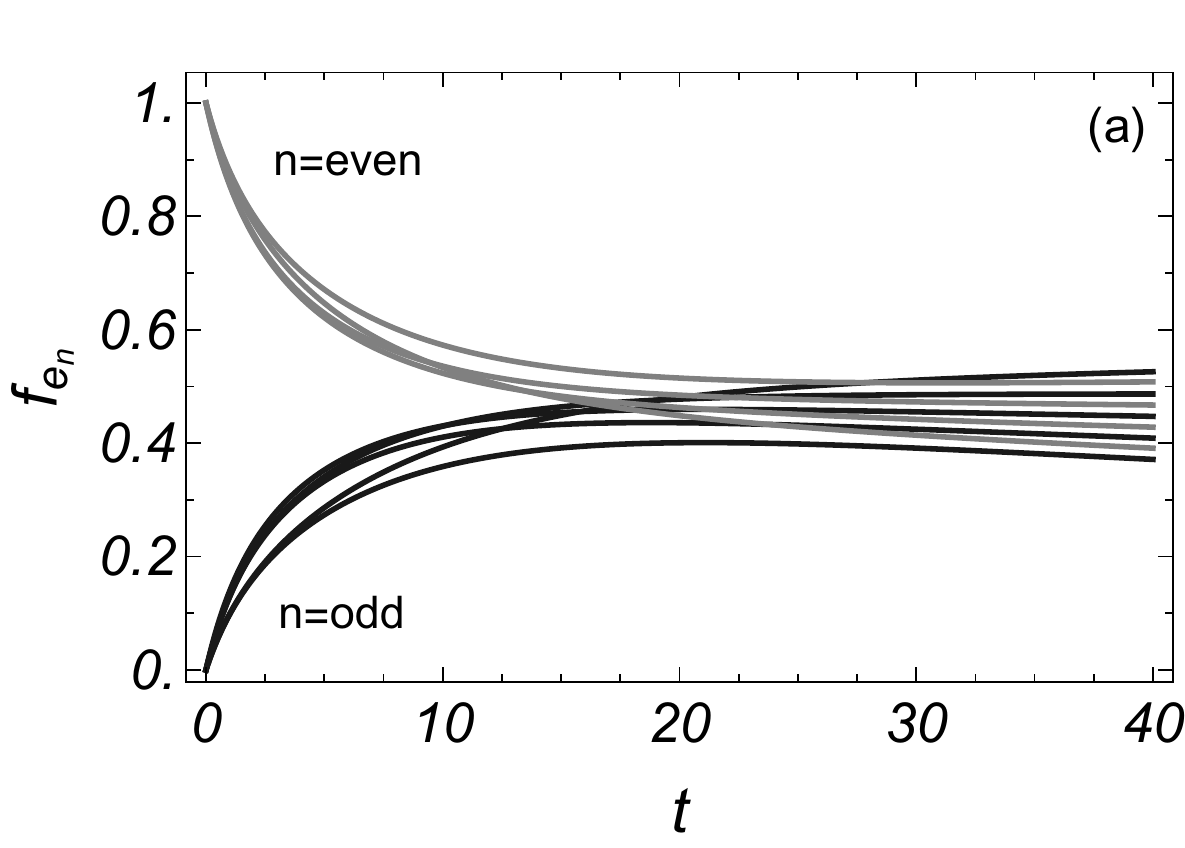}\\
\includegraphics[width=.83\linewidth]{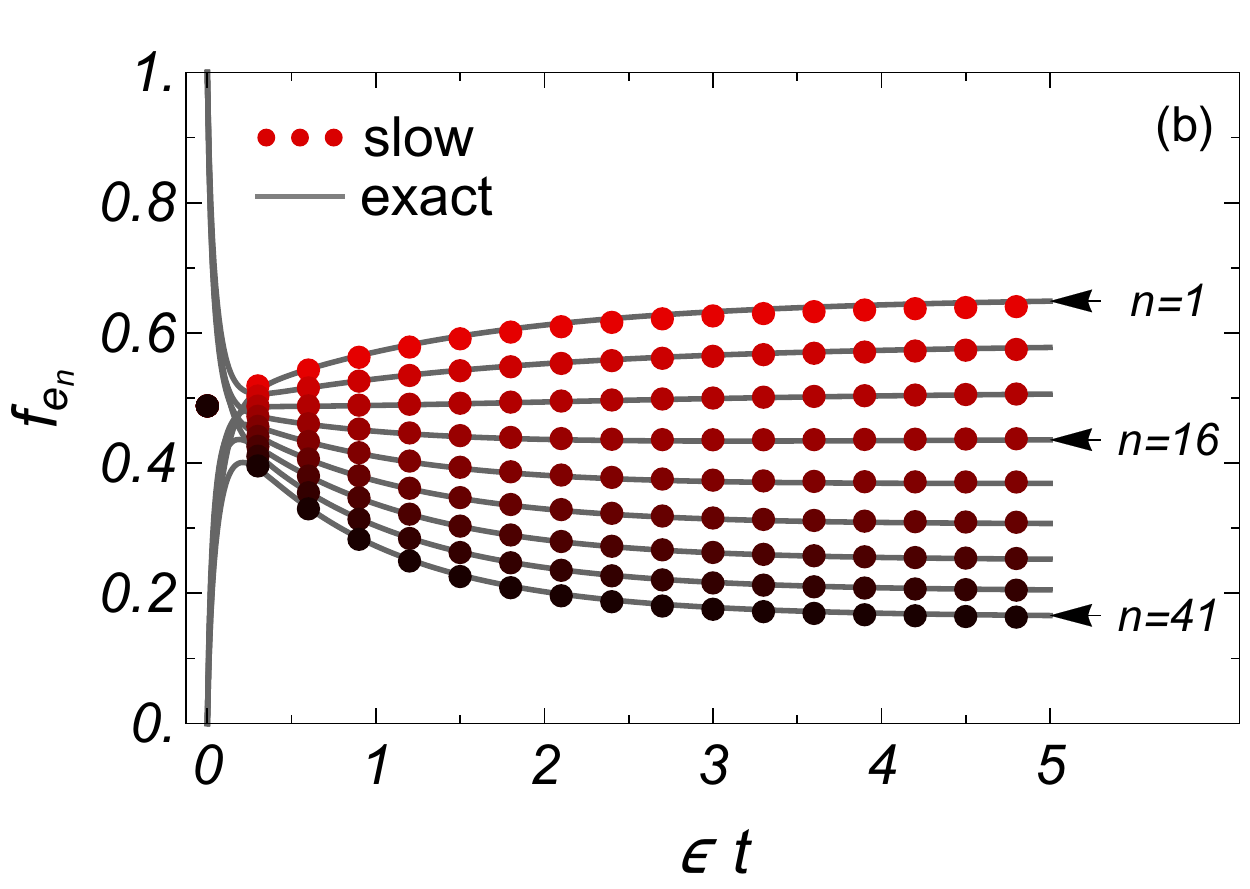}
\caption{(Color online) Time evolution of the occupation function $f_{e_n}$ shown for $n=1,6,11,\dots, 41$ ($L=41$) and 
$\epsilon=0.01$ starting from an initial state with $f_{e_n}=1$ ($f_{e_n}=0$) for states with even (odd) $n$, respectively. (a)  On short time scales the system relaxes towards a state with $\beta\approx 0$ and equal occupation of all levels. (b) The time evolution toward the steady state occurs on a time scale set by $1/\epsilon$ and therefore the time axis has been rescaled by a factor $\epsilon$. The points are obtained by solving the time evolution of the Lagrange parameters using Eq.~\eqref{EqLambdaProp} which then determine a  Fermi distribution function. The comparison with the exact solution of the Boltzmann equation (lines) shows that this allows for a quantitative description of the slow dynamics for small $\epsilon$. \label{figProp}}
\end{figure}

{\it Projection operator ---} 
The two slow modes $q_i$, corresponding to $\partial\rho_0/\partial \lambda_i$ in the main text, are defined as 
\begin{align}
&q_1^{\beta,\mu}(e)=\frac{\partial f^0_e}{\partial \beta}, \quad
q_2^{\beta,\mu}(e)=\frac{\partial f^0_e}{\partial(-\beta\mu)}
\end{align}
where we identified $\lambda_2=-\beta \mu$.
Denoting the conserved quantities by
\begin{equation}
c_1(e)=e, \
c_2(e)=1
\end{equation}
the analog of the  projector super-operator defined by  Eq.~(\ref{EqP}) is simply given by
\begin{align}\label{EqPBolz}
&\hat{P}[X]= -\sum_{i,j=1}^2 q_i ^{\beta,\mu}(e) \ (\chi^{-1})_{ij}  \int de\, (c_j(e)  X(e)) \\ 
&\chi_{ij}(\beta,\mu)=-\int de \, c_i(e) \,q_j ^{\beta,\mu}(e) .\notag 
\end{align}
With these notations, the steady state condition, Eq.~(\ref{EqBolz0}), is the analog of Eq.~(\ref{EqRho0})
$$ 
\int de   \,  c_i(e)  D[f^0]_e=0
\quad \leftrightarrow \quad
\tr[C_i \cl{L}_1\rho_0]=0
$$
or, equivalently, 
$$\hat{P}\big[D[f^0]\big]=0\quad \leftrightarrow \quad
\hat P (\cl{L}_1\rho_0)=0. $$

\begin{figure}[b!]
\includegraphics[width=.65\linewidth]{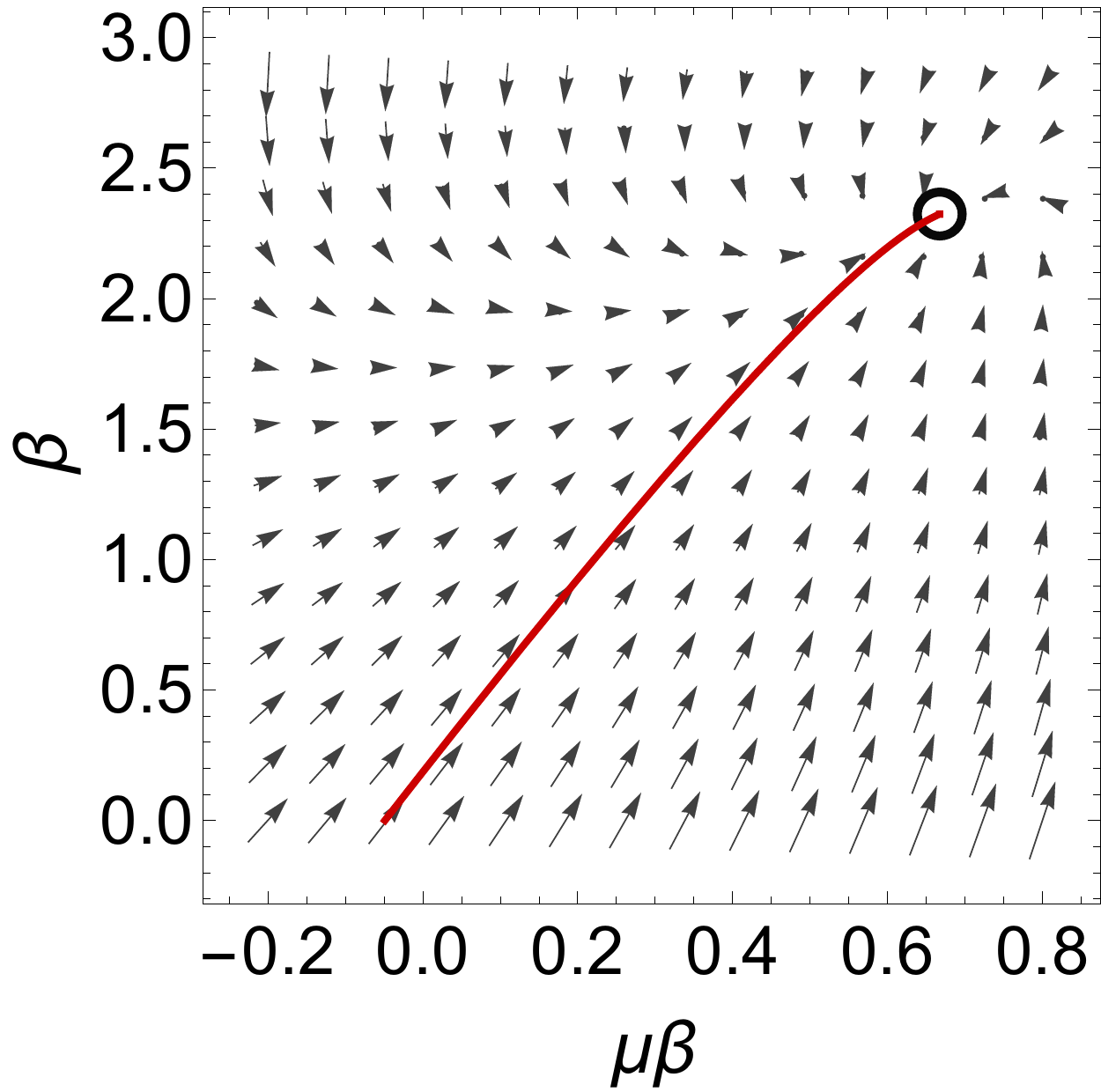}
\caption{(Color online) 
Force fields calculated using Eq.~\eqref{EqForceBolz} shown for $L=41$. The force fields determine the time evolution of the Lagrange parameters according to Eq.~\eqref{EqLambdaProp}. The red solid line shows the trajectory for the initial conditions 
used in Fig.~\ref{figProp}, where a comparision to the exact solution of the Boltzmann equation is shown. The unique stationary state is indicated by  a black circle.
 \label{figF}}
\end{figure}

{\it Relaxation towards the steady state ---} 
While the main focus of this paper is the computation of the steady state, we briefly discuss the relaxation towards the steady state, which can easily be computed exactly by solving the Boltzmann equation.

Similarly as GGEs are expected to be a fairly good description shortly after the system has prethermalized
\cite{berges04,Aarts2000,Gring2012,Langen2015,Langen2016,moeckel08,Kollath2007,kollar11,Marino2012,Mitra2013,Marcuzzi2013,bertini15,Demler2015,Buchhold2014,Marcuzzi2016,Barnett2011,NessiIucci2014,Nowak2014,canovi16,golez16,Chiocchetta2017}
to the GGE manifold, also here the Fermi-Dirac distribution describes approximately not only the steady state but also the approach to it. After a few collisions level occupations can be approximately described by a Fermi distribution with time-dependent parameters $\{\beta(t),\mu(t)\}$, whose evolution is determined by the forces introduced in Eq.~(\ref{EqForce}).

In this case force field, see Eq.~\eqref{EqForce}, has only two components,
\begin{equation}\label{EqForceBolz}
F_i(\beta,\mu)= -\big(\chi^{-1}\big)_{ij} \int de\, c_j(e)  \,  \epsilon D[f^0]_e
\end{equation}
and can be used to propagate parameters $\beta$ and $\mu$, 
\begin{equation}\label{EqLambdaProp}
\frac{d\beta}{dt}(t)=F_1(t), \quad
\frac{d(-\beta\mu)}{dt}(t)=F_2(t),
\end{equation}
as soon as our system is approximately described by a Fermi-Dirac distribution. These equations are only valid to leading order in $\epsilon$.

Fig.~\ref{figProp}  shows the  time-evolution obtained from a numerical solution of the Boltzmann equation for $\epsilon=0.01$.
We initialize our system with the non-equilibrium state $f_{e_n}=1$ for states with even and  $f_{e_n}=0$ for states with odd $n$. Collisions lead to a rapid relaxation of this non-equilibrium state to a thermal state. We find that within our model this time scale is $\tau_0\sim  5$, independent of $\epsilon$ for small $\epsilon$. As the collision processes conserve energy and particle number, the approximate thermal state is determined by the initial values of energy and particle number, which leads to an initial relaxation towards a state with 
 $\beta=0$ and $\beta \mu=0$ (in the thermodynamic limit).
The subsequent dynamics is driven by the processes violating particle number and energy conservation (the gain and loss terms)  and occurs on a time scale set by $1/\epsilon$.  We find that these processes are quantitatively described by Eq.~(\ref{EqLambdaProp}), as can be seen by comparing the dots and the solid line in the lower panel of Fig.~\ref{figProp}.

Fig.~\ref{figF} shows the force fields calculated according to Eq.~\eqref{EqForceBolz} and the resulting dynamics in the space of Lagrange parameters.

\begin{figure}[!t]
\center \includegraphics[width=.83\linewidth]{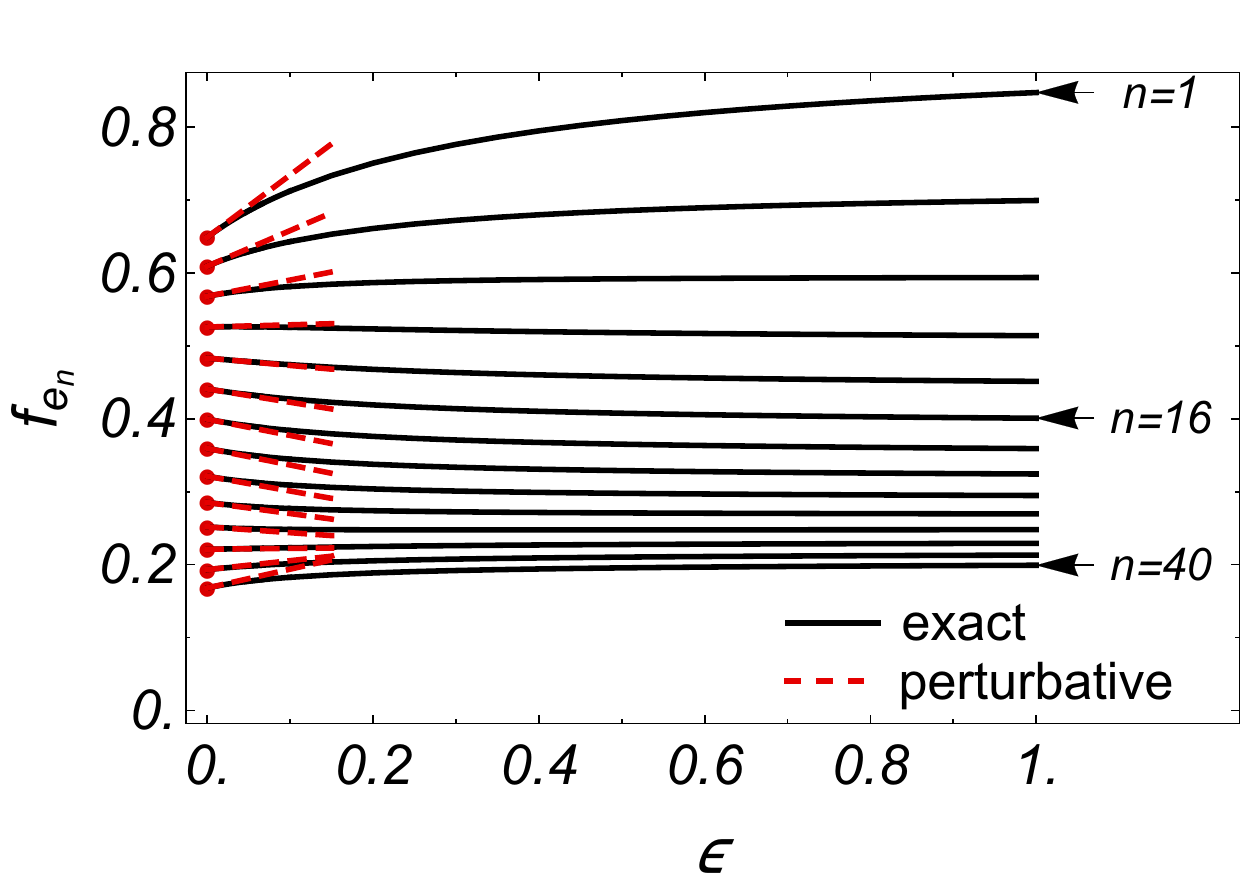}
\caption{(Color online) Level occupation $f_{e_n}$ as a function of perturbation strength $\epsilon$. Solid lines are obtained from the exact calculation using Boltzmann equation, Eq.~\eqref{EqBoltzmann}, while dashed lines are obtained from our perturbative approach, including zeroth and first order in $\epsilon$. Only every third $n$ is shown for system with $L=41$ single-particle states.\label{figFnEps}}
\end{figure}

\begin{figure}[!t]
\center \includegraphics[width=.83\linewidth]{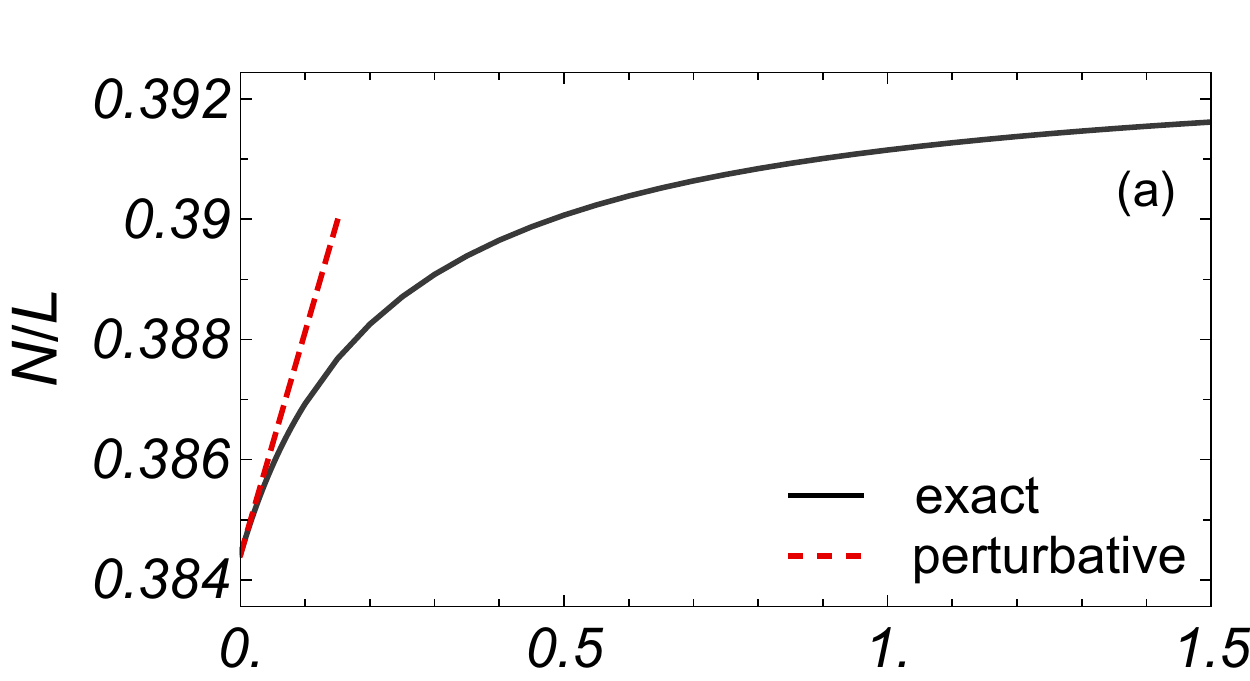}\\
\vspace{-0.6cm}
\center \includegraphics[width=.83\linewidth]{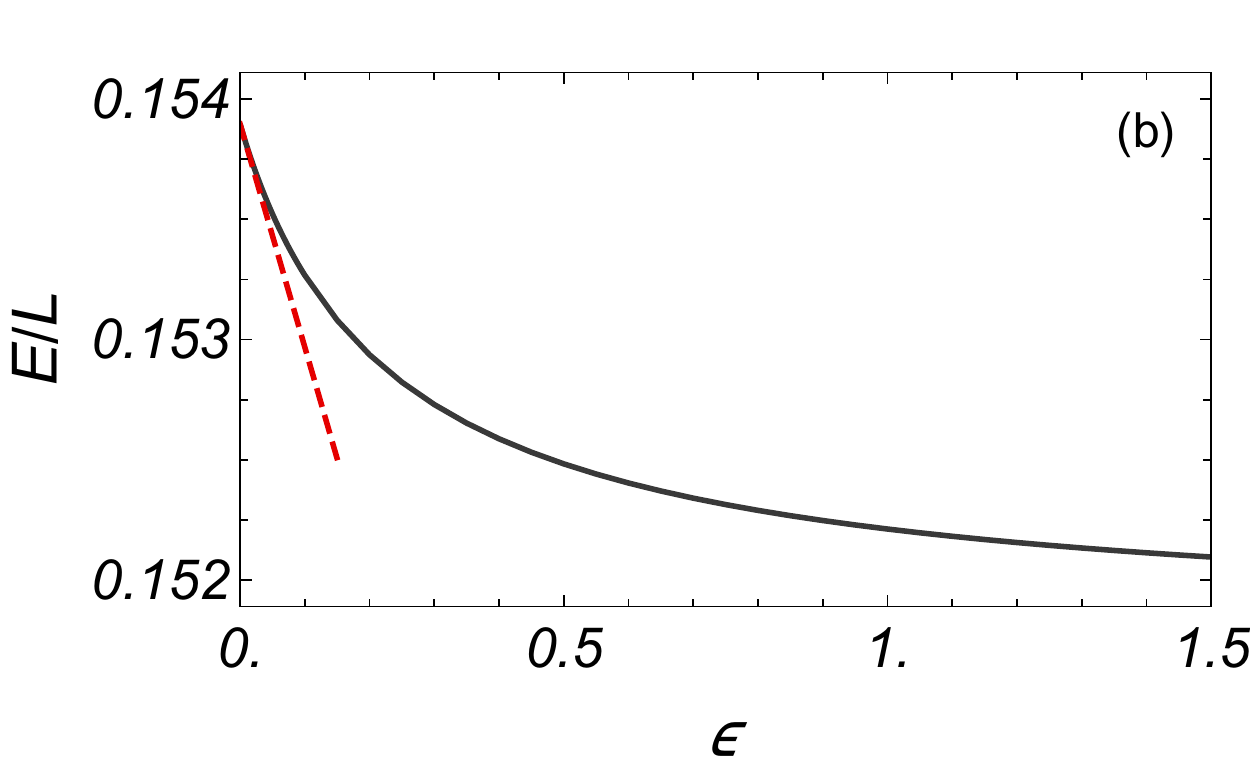}
\caption{(Color online) (a) Particle and (b) energy density as a function of perturbation strength $\epsilon$. The dashed line shows the result of perturbation theory, including zeroth and first order in $\epsilon$. \label{figNE}}
\end{figure}

{\it Perturbation theory for the steady state ---}
As a next step, we compare the exact steady-state solution of the Boltzmann equation for finite $\epsilon$  to the result of perturbation theory to linear order in $\epsilon$.
We expand around the Fermi functions $f^0_{e}(\beta,\mu)$ with parameters $\beta$ and $\mu$ obtained from the `zeroth order' described above. To linear order in $\epsilon$, we obtain
using $f_e=f^0_e + \epsilon f^1_e$ to $\mathcal{O}(\epsilon)$
\begin{align}
0&=\epsilon(M^{(0)} + \epsilon D^{(1)}) [f^{1}] + \epsilon D[f^0]\label{EqBoltz1st}\\
\Leftrightarrow f^1&=-(M^{(0)} + \epsilon D^{(1)})^{-1}  D[f^0] \notag
\end{align}
which has precisely the same form as Eq.~\eqref{EqDeltaRho} in the Liouville case.
For the discrete case matrices $M^{(0)},D^{(1)}$ are defined using a straightforward Taylor expansion
\begin{align}\label{EqLinApprox}
M[f]_{e_n}&\approx 
\epsilon \sum_{n'}M^{(0)}_{n,n'} f^{1}_{e_{n'}}
\\
(-l_{e_n} f_{e_n} + g_{e_n} \bar{f}_{e_n}) &\approx 
D[f^0]_{e_n}+\epsilon \sum_{n'} D_{n,n'}^{(1)} f^{1}_{e_n'}. \notag  
\end{align}
Note that $D^{(1)}$ is a diagonal matrix in our example. 
To obtain $\boldsymbol{f}^1=\{f^1_{e_1},f^1_{e_2}, \dots, f^{1}_{e_L}\}$ corrections we can use the perturbation theory developed in Sec.~\ref{SecMarkPert}.
As before we need to treat the ($\parallel$) and ($\perp$) subspaces separately.
Following Eqs.~(\ref{EqLinRespP},\ref{EqLinRespQ}),
\begin{align}
&\boldsymbol{f}^1=\boldsymbol{f}^1_{\parallel} + \boldsymbol{f}^1_{\perp}, \label{EqLinBoltz}\\
&\boldsymbol{f}^1_{\perp}=-(QM^{(0)}Q)^{-1} \ \boldsymbol{D}[\boldsymbol{f}^0] \notag\\
&\boldsymbol{f}^1_{\parallel}=(PD^{(1)}P)^{-1} \ PD^{(1)} Q \ (QM^{(0)}Q)^{-1} \ \boldsymbol{D}[\boldsymbol{f}^0],\notag
\end{align} 
where $P, Q$ are written as matrices after evaluating  Eq.~\eqref{EqPBolz} for discretized energies and  $\hat{Q}=\hat{1}-\hat{P}$. 
Fig.~\ref{figFnEps} shows steady state level occupation $f_{e_n}$ as a function of perturbation strength $\epsilon$. In the limit $\epsilon\to 0$ Fermi functions obtained from condition Eq.~\eqref{EqBolz0} give the exact steady state occupation distributions. For finite $\epsilon \lesssim 0.05$ the corrections linear in $\epsilon$ describe well the exact result. In Fig.~\ref{figNE} we show similar results for the total energy and particle number of the system, which for the chosen model change only slightly.

It is also straightforward to calculate higher order corrections using perturbation theory applied to the Boltzmann equation. However, here one has to take into account that the Boltzmann equation is a non-linear equation in the occupation functions $f_e$ while the Liouville equation is a linear equation in the density matrix $\rho$. Therefore the equations used to obtain higher order corrections differ in the two cases but the overall structure of perturbation theory (separation into parallel and perpendicular sector) remains the same.

\section{Conclusions}

The main observation, which is the basis of our study, is that many driven systems can be approximately but efficiently described by generalized Gibbs ensembles built from approximately conserved quantities. Such approximate conservation laws are important for many different systems. For example, in almost all solids, the coupling of electrons and phonons is weak due to the large mismatch of ionic and electronic masses. Therefore the difference of the electron and phonon Hamiltonians, $H_e-H_{\rm ph}$, is an approximate conservation law widely used in two-temperature models \cite{anisimov74,armstrong81,eesley86,allen87,singh10} which efficiently describe non-equilibrium states of solids.
A more exotic example is the description of spin-chain materials using as the approximate conservation laws those of the underlying integrable Heisenberg model \cite{lange17}. For such situations, we developed in this paper a perturbation theory for the steady state by perturbing around generalized Gibbs ensembles. Here it was essential to treat perturbations perpendicular and parallel to the manifold of generalized Gibbs states separately.

To show the validity of our concepts we studied a small finite size quantum system with Lindblad dynamics. For the future \cite{inpreparation} we plan to test numerically our approach also for larger quantum systems close to the thermodynamic limit. In this paper we considered as an example for a large system only a model described by the Boltzmann equation.
From a more general point of view, one can consider the Boltzmann equation also as a rate equation for approximately conserved quantities, the quasi-particle occupations $c^\dagger_k c_k$, thus fixing the parameters of the appropriate generalized Gibbs ensemble $e^{-\sum_k g_k c^\dagger_k c_k}$ (as in our paper, we assume a translationally invariant system here). Therefore it would be interesting to use our formalism to calculate corrections to the non-equilibrium steady state not captured by the Boltzmann approach. A comparison of the two formalisms and a systematic calculation of beyond-Boltzmann corrections for weakly interacting quantum systems is left for further studies. In the classical context such a formalism has been developed by Gurarie \cite{gurarie95}.
\comm{It would be in general interesting to test the validity of our approach in purely classical systems, for example, those described by the Fokker-Planck equation.}

An obvious question is whether our perturbation theory in $\epsilon$, where $\epsilon$ is the prefactor of driving terms breaking the conservation laws, has a finite radius of convergence and/or whether all expansion coefficients are finite. Similarly to many other (highly successful) perturbative expansions in physics, the radius of convergence is formally expected to be zero. This is obvious in the case when the perturbation arises from a Lindblad operator, where negative $\epsilon$ correspond to negative friction and therefore an unstable situation for arbitrarily small $\epsilon$, which implies a vanishing radius of convergence. The expansion in $\epsilon$ should therefore be viewed as an asymptotic expansion similar to, e.g., the expansion in the fine-structure constant $\alpha$ in quantum-electrodynamics.
A more difficult question is whether one can expect that all expansion coefficient in powers of $\epsilon$ are finite or not. This is an open question which requires further studies. There is, however, a possible mechanism which can induce non-analytic corrections: in the presence of exact symmetries, hydrodynamic modes exist which enforce long-ranged correlations and, for example, the presence of long-time tails after quenches \cite{lux14,bohrdt17,leviatan17}. These hydrodynamic modes obtain masses proportional to $\epsilon$ or $\epsilon^2$ when the relevant conservation laws are weakly
violated. A perturbative expansion in these masses can lead to non-analytic corrections even when perturbing around a GGE.

For the future it will be interesting to generalize our perturbative expansion for the steady state to dynamical questions. For example, we have already shown for a simple example based on the  Boltzmann equation that time-dependent generalized Gibbs ensembles are a good starting point to investigate the approach of the steady state. It will be interesting to develop a perturbative expansion for such situations which will also allow to describe the dynamical response of driven systems and/or situations where instead of a unique steady state a limit cycle (time crystal) is realized.

\acknowledgments{
We acknowledge useful discussions with M. Genske, J. Lux, M. \v{Z}nidari\v{c} and financial support of the German Science Foundation under CRC 1238 (project C04).
}

\appendix{} \label{App}
\section{Regularization of inverse}\label{AppRegularize}
Since $\cl{L}$ has a zero mode, $\cl{L}^{-1}$ is singular therefore formally one has to regularize it as in Eq.~(\ref{EqDeltaRho}), which at the same time also avoids solution $\delta\rho=-\rho_0$ that satisfies $\cl{L}(\rho_0+\delta\rho)$ trivially. If we write 
$\rho_0=\sum_{\alpha}c_\alpha\rho^\alpha$  
in terms of $\cl{L}$ (right) eigenstates $\rho^\alpha$, 
$\cl{L}\rho^\alpha=\lambda_\alpha\rho^\alpha$,
then the regularization 
\begin{align}
\delta\rho
&=-\lim_{\eta\to 0}(\cl{L}-\eta \mathbb{1})^{-1}\cl{L}_1\rho_0
=-\lim_{\eta\to 0}(\cl{L}-\eta \mathbb{1})^{-1}\cl{L}\rho_0,\notag\\
&=-\lim_{\eta\to 0}(\cl{L}-\eta \mathbb{1})^{-1} \ \sum_{\alpha}\lambda_\alpha c_\alpha\rho^\alpha\notag\\
&=-\lim_{\eta\to 0}\sum_{\alpha} \frac{\lambda_\alpha}{\lambda_\alpha - \eta} \ c_\alpha\rho^\alpha 
=-\sum_{\alpha, \lambda_\alpha\neq 0} c_\alpha\rho^\alpha . \label{drhoReg}
\end{align}
gives  the correct result, $\rho_0+\delta \rho = \sum_{\alpha, \lambda_\alpha=0} \rho^\alpha$.
The sign $\eta>0$ is obtained from the property that $\text{Re}\, \lambda_\alpha \le 0$ guaranteeing the absence of exponentially growing solutions. 

The regularization also guarantees that $\tr \, \delta \rho=0$. From $\tr \, \dot \rho = \tr[\cl{L} \rho]=0$, it follows that 
 $\lambda_\alpha \tr\,\rho^\alpha=0$ and therefore $\tr\,\rho^\alpha=0$
for all $\alpha$ with $\lambda_\alpha\neq 0$. Using  \eqref{drhoReg}, we therefore find $\tr\, \delta \rho=0$.

\section{Identities for super-operators}\label{AppProj}
Below we check the properties of super-projector defined in  Eq.~(\ref{EqP}).\\
\\
\noindent{\em Projection property}
\begin{align}
\hat{P}^2 X
&=\sum_{i'j'ij} \frac{\partial\rho_0}{\partial\lambda_{i'}} (\chi^{-1})_{i'j'} \ 
\tr\big[C_{j'}\frac{\partial\rho_0}{\partial\lambda_i}\big] (\chi^{-1})_{ij} \ \tr[C_jX]\notag\\
&=-\sum_{i'j'ij} \frac{\partial\rho_0}{\partial\lambda_{i'}} (\chi^{-1})_{i'j'} \ 
\chi_{j'i} (\chi^{-1})_{ij} \ \tr[C_jX]\notag\\
&=-\sum_{i'j'j} \frac{\partial\rho_0}{\partial\lambda_{i'}} (\chi^{-1})_{i'j'} \ 
\delta_{j'j} \ \tr[C_jX]
=\hat{P} X
\end{align}
%

\noindent{\em Inverse within $P$ subspace}\\
Since $(\hat{P}\cl{L}_1\hat{P})^{-1}$ is essential for building up the perturbation theory we confirm  the validity of Eq.~(\ref{EqPLPinv})
\begin{align}\label{EqPinv2}
(\hat{P}\cl{L}_1\hat{P})^{-1} \hat{P} X
&=-\sum_{pr}\frac{\partial\rho_0}{\partial\lambda_p} (M^{-1})_{pr} \ \tr[C_r X],
\end{align}
with 
$M_{pr}
=-\tr[C_p \cl{L}_1 \frac{\partial\rho_0}{\partial\lambda_r}]
=\langle \dot{C}_p C_r \rangle_{_{0,c}}$ 
by calculating
\begin{align}
&(\hat{P}\cl{L}_1\hat{P})^{-1}(\hat{P}\cl{L}_1\hat{P})X\notag\\
&=-\sum_{pr,kl,ij}\frac{\partial \rho_0}{\partial \lambda_p} 
(M^{-1})_{pr} \
\chi_{rk}
(\chi^{-1})_{kl} M_{li} \ (\chi^{-1})_{ij} \tr[C_j X]\notag\\
&=-\sum_{pr,kl,ij}\frac{\partial \rho_0}{\partial \lambda_p} (M^{-1})_{pr} \
\delta_{r,l} M_{li} \ (\chi^{-1})_{ij} \tr[C_j X]
=\hat{P}X.\notag
\end{align}
%

\comm{
\section{Properties of projectors and inversion of projected Lindblad operators}\label{AppProj2}
In this appendix we discuss the relation of the projectors $\hat P $ and $\tilde P$ and describe how projected Lindblad operators can be inverted.

We first note that any projected density matrix is traceless
\begin{equation}
\tr[\hat P \rho]=0
\end{equation}
which follows from $\tr[\rho_0]=1$ and therefore $\partial \tr[\rho_0]/\partial \lambda_i=0$.

If we consider all conservation laws from the set $\mathcal Q=\{ |n\rangle \langle m|\  {\rm with}\  E_n^0=E_m^0\}$, $H_0\ket{n}=E_n^0 \ket{n}$, then we obtain
\begin{equation}
\hat{P}X= \tilde P  X- \frac{\tr[X]}{\tr[\mathbb{1}]	} \, \mathbb{1}	
\end{equation}
where $ \tilde P X$ is part of $X$ that can be written in terms of elements of $\mathcal{Q}$ (for non-degenerate case simply the diagonal part). Therefore $\hat P$ projects on conservation laws and subtracts the trace.

Assuming that $\hat{P} \cl{L}_1 \rho_0$, Eq.~(\ref{rho0Pro1}),
 has a unique solution with trace 1, then one can easily show that $\hat{P} \cl{L}_1 \hat{P}$ is invertible in $\hat P$ space. Assuming that $\hat{P} \cl{L}_1 \hat{P} \rho_1=0$, it follows immediately that $\hat{P} \cl{L}_1 (\rho_0+\hat P \rho_1)=0$. As $\rho_0$ is by assumption a unique solution, we find $\hat P \rho_1=0$. Therefore $\rho_1$ has no component in $\hat P$ space.

As $\hat P \mathbb{1} =0$, the identity matrix is in the $\hat Q$ space, $\hat Q \mathbb{1}=\mathbb{1}$. This seems to be a problem as we have to invert $\hat Q \cl{L}_0 \hat Q$ and  
$\cl{L}_0 \mathbb{1}=0$ for all $\cl{L}_0$ describing unitary evolution . In all of our formulas, however, the inverse of $\hat Q \cl{L}_0 \hat Q$ is only applied to {\em traceless} density matrices, therefore no singularity arises from this zero mode. For numerical implementations one can simply add a 'mass term' $m_0$ to this zero mode by replacing $\cl{L}_0 \to \cl{L}_0+ m_0  \hat P_\mathbb{1}$  where $\hat P_\mathbb{1}$ is the superoperator projecting onto the identity matrix  defined by $\hat P_\mathbb{1} X = \mathbb{1} \, \tr[X]/\tr[\mathbb{1}]$.

}

\comm{
\section{Correlation functions}\label{AppCorr}
Higher order corrections $\delta\rho$ to $\rho_0$, as derived from our perturbation theory, can also be expressed in terms of multi-time correlation functions. In the main text we have already introduced
\begin{align}\label{EqCorFunc00}
\chi_{ij}&=-\tr[C_i \partial \rho_0/\partial \lambda_j]
	=\ave{C_i C_j}_{0,c}\\
M_{ij}&=-\tr[C_i \cl{L}_1 (\rho_0/\partial \lambda_j)]=\ave{\dot{C_i} C_j}_{0,c}\notag
\end{align}
however, to express higher order $\mathcal{O}(\epsilon^n)$ contributions one needs also more complicated correlation functions
\begin{align}\label{EqCorFunc0}
N^{(n)}_i=&\tr[C_i \cl{L}_1 (\cl{L}_0^{-1} \cl{L}_1)^n \rho_0]=
\tr[\dot{C}_i (\cl{L}_0^{-1} \cl{L}_1)^n \rho_0]\\
M^{(n)}_{ij}=&-\tr[\dot{C}_i (\cl{L}_0^{-1} \cl{L}_1)^n (\partial \rho_0/\partial \lambda_j)]\notag
\end{align}
In order to write them in a compact form with an explicit time ordering we introduce a general notation of {\it Lindblad super-operators in the interacting picture},
\begin{align}
&\cl{L}_1(t)\rho
=  L_\alpha(t) \rho L_\alpha^\dagger(t) 
- \frac{1}{2} \{L_\alpha^\dagger(t) L_\alpha(t), \rho\}
\end{align} 
where Lindblad operators $L_\alpha(t)=e^{iH_0 t} \, L_\alpha \, e^{-iH_0t}$ are evolved with respect to $H_0$.\\\\
}
\begin{widetext}
\comm{
Using this notation the lowest order correlation functions can be, according to Eq.~\eqref{drhoReg}, written as regularized time integrals
\begin{align}
N_{j}^{(1)}
&=\tr\big[C_j\cl{L}_1 (\cl{L}_0^{-1} \cl{L}_1)\rho_0\big]
=-\int_{-\infty}^{\infty}   dt_1 \, \theta(t_1) e^{-\eta t_1} \, \tr\big[ \dot{C}_j(t_1) \cl{L}_1(0) \rho_0 \big] \label{EqCorFuncNj1}\\
N_{j}^{(2)}
&=\tr\big[C_j\cl{L}_1 (\cl{L}_0^{-1} \cl{L}_1)^2\rho_0\big] 
=\int_{-\infty}^{\infty}  dt_1 dt_2\, \theta(t_1)\theta(t_2) e^{-\eta (t_1+t_2)}
\tr\big[ \dot{C}_j(t_1+t_2) \cl{L}_1(t_1)\cl{L}_1(0) \rho_0 \big]\\
M_{jk}^{(1)}
&=-\tr\big[C_j\cl{L}_1 (\cl{L}_0^{-1} \cl{L}_1)\frac{\partial \rho_0}{\partial \lambda_k}\big]
=\int_{-\infty}^{\infty}  dt_1 \, \theta(t_1) e^{-\eta t_1} \, \tr\big[ \dot{C}_j(t_1) \cl{L}_1(0) \frac{\partial \rho_0}{\partial \lambda_k} \big]\\
M_{jk}^{(2)}
&=-\tr\big[C_j\cl{L}_1 (\cl{L}_0^{-1} \cl{L}_1)^2\frac{\partial \rho_0}{\partial \lambda_k}\big] 
=-\int_{-\infty}^{\infty}   dt_1 dt_2\, \theta(t_1)\theta(t_2) e^{-\eta (t_1+t_2)}
\tr\big[ \dot{C}_j(t_1+t_2) \cl{L}_1(t_1)\cl{L}_1(0) \frac{\partial \rho_0}{\partial \lambda_k} \big]
\end{align}
Note that the time ordering of (super-)operators is such that it is most naturally represented using Keldysh formalism. \\\\
{\it First order}\\
Using the correlation functions defined above and the equality 
$\tr\big[C_k\cl{L}_1 \hat{Q}\cl{L}_0^{-1} \hat{Q}\cl{L}_1\rho_0\big]=\tr\big[C_k\cl{L}_1\cl{L}_0^{-1}\cl{L}_1\rho_0\big]$, following from $\hat{P}\cl{L}_1\rho_0=0$ for properly chosen $\rho_0$, the linear contribution to the expectation values of conserved quantities, Eq.~\eqref{EqLinRespC}, gets the form
\begin{equation}
\ave{C_i}_1
=\sum_{jk} \chi_{ij} \, (M^{-1})_{jk} 
\, \tr\big[C_k\cl{L}_1 \hat{Q}\cl{L}_0^{-1} \hat{Q}\cl{L}_1\rho_0\big]
=\sum_{jk} \chi_{ij} \, (M^{-1})_{jk} N^{(1)}_k
\end{equation}
{\it Second order}\\
Second order $\mathcal{O}(\epsilon^2)$ in the tangential part of $\delta\rho$ (relevant for expectation values of conserved quantities) has two contributions, 
$\delta\rho_{2,\parallel}=\delta\rho^{(1)}_{2,\parallel}+\delta\rho^{(2)}_{2,\parallel}$
\begin{align}
\delta\rho^{(1)}_{2,\parallel}]
=&-(\hat{P}\cl{L}_1 \hat{P})^{-1}\, \hat{P}\cl{L}_1 \hat{Q} \cl{L}_0^{-1} \hat{Q}\cl{L}_1 \hat{Q}\cl{L}_0^{-1}\hat{Q}\cl{L}_1 \rho_0\notag\\
=&-(\hat{P}\cl{L}_1 \hat{P})^{-1} \, \hat{P}\cl{L}_1 \cl{L}_0^{-1} \cl{L}_1 \cl{L}_0^{-1}\cl{L}_1 \rho_0
+ \hat{P} \cl{L}_0^{-1} \hat{P} \cl{L}_1 \cl{L}_0^{-1}\cl{L}_1 \rho_0\\
\delta\rho^{(2)}_{2,\parallel}
=&(\hat{P}\cl{L}_1 \hat{P})^{-1} \, \hat{P}\cl{L}_1 \hat{Q} \, \cl{L}_0^{-1} \,
 \hat{Q}\cl{L}_1 \hat{P} \ (\hat{P}\cl{L}_1 \hat{P})^{-1} \hat{P}\cl{L}_1 \hat{Q}\cl{L}_0^{-1}\hat{Q}\cl{L}_1 \rho_0 \notag\\
=&(\hat{P}\cl{L}_1 \hat{P})^{-1} \, \hat{P}\cl{L}_1 \cl{L}_0^{-1}  
\cl{L}_1 \hat{P} \, (\hat{P}\cl{L}_1 \hat{P})^{-1} \hat{P}\cl{L}_1 \cl{L}_0^{-1}\cl{L}_1 \rho_0 
-\hat{P} \cl{L}_0^{-1} \hat{P} \cl{L}_1 \cl{L}_0^{-1}\cl{L}_1 \rho_0
\end{align}
which we obtain by inserting $\hat{Q}=\hat{1}-\hat{P}$ under condition $\hat{P}\cl{L}_1\rho_0=0$ and using that $\cl{L}_0^{-1}$ cannot change the subspace, i.e., $\hat{P}\cl{L}_0^{-1}=\cl{L}_0^{-1}\hat{P}=\hat{P}\cl{L}_0^{-1}\hat{P}$. The two contributions then add up to
\begin{align}
\delta\rho_{2,\parallel}
=&-(\hat{P}\cl{L}_1 \hat{P})^{-1} \, \hat{P}\cl{L}_1 \cl{L}_0^{-1} \cl{L}_1 \cl{L}_0^{-1}\cl{L}_1 \rho_0
+(\hat{P}\cl{L}_1 \hat{P})^{-1} \, \hat{P}\cl{L}_1 \cl{L}_0^{-1}  
\cl{L}_1 \hat{P} \, (\hat{P}\cl{L}_1 \hat{P})^{-1} \hat{P}\cl{L}_1 \cl{L}_0^{-1}\cl{L}_1 \rho_0
\end{align}
Using this expression it is straightforward to write the $\mathcal{O}(\epsilon^2)$ contribution to the expectation value of conserved quantities in the steady state in terms of correlation functions. One should notice that each $\hat{P}$ 'cuts' the expression in the sense that content between to consequential $\hat{P}$ corresponds to one correlation function. The inverses $(\hat{P} \cl{L}_1 \hat{P})^{-1}$ are treated as in Eq.~\eqref{EqPinv2}, contribute $M^{-1}$ components to the final expression. Using the definition of projector $\hat{P}$, Eq.~\eqref{EqP}, the definition of the inverse $(\hat{P} \cl{L}_1 \hat{P})^{-1}$, Eq.~\eqref{EqPinv2}, and definitions of multi-time correlation functions, Eq.~\eqref{EqCorFunc0}, one obtains the following expression
\begin{align}
\ave{C_i}_2
=\tr[C_i \delta\rho_{2,\parallel}]
=&-\sum_{j,k} \chi_{ij} \, (M^{-1})_{jk}\,  N^{(2)}_k 
+ \sum_{j,k,l,m} \chi_{ij} \, (M^{-1})_{jk} \,  M^{(1)}_{kl} \, (M^{-1})_{lm} \, N^{(1)}_m.
\end{align}
}

\noindent{\em Lehmann representation}\\
All expressions obtained from our perturbation theory can be calculated in a straightforward way using eigenstates and eigenenergies of $H_0$.  As Lehmann representations are perhaps less common in the case of a Lindblad dynamics we show, for example, the Lehmann representation of $N_j^{(1)}$, Eq.~\eqref{EqCorFuncNj1}, needed to calculate $\ave{C_i}_1$. By inserting the identity operator $\sum_m |m\rangle \langle m|$
between all operators and super-operators one obtains
\begin{align}
N_j^{(1)}=
\sum_{mnpr}
\sum_{\alpha\beta} \frac{i}{(E_n^0-E_p^0-i\eta)}
\, \big(C^{(j)}_{r} L^{(\beta)}_{rn} L^{(\beta)\dagger}_{pr} -
\frac{1}{2}(C^{(j)}_{p} + C^{(j)}_{n}) L^{(\beta)\dagger}_{pr} L^{(\beta)}_{rn}\big)
\, \big(L^{(\alpha)}_{nm} L^{(\alpha)\dagger}_{mp}\rho_{m}^{0} 
- \frac{1}{2}(\rho_{p}^{0}+\rho_{n}^{0})L^{(\alpha)\dagger}_{nm} L^{(\alpha)}_{mp}\big), \label{EqTrLehman}
\end{align}
where $m,n,p,r$ run over the eigenstates $\ket{m}$, 
$H_0\ket{m}=E_m^0\ket{m}$, while $\alpha,\beta$ run over the Lindblad operators defined in Eq.~\eqref{lindblad}.
We use notation 
$\ave{m|L_\alpha|n}=L^{(\alpha)}_{mn}$ and 
$\ave{m|\rho_0|m}=\rho_{m}^{0}$, $\ave{m|C_j|m}=C^{(j)}_{m}$ where we assumed that the conserved quantities $C_i$ are diagonal operators in the eigenbasis of $H_0$. As $\cl{L}_0 |n\rangle \langle m|=-i (E_n^0-E_m^0)  |n\rangle \langle m|$, the inverse is obtained as $\cl{L}_0^{-1} |n\rangle \langle m|
=\frac{i}{E_n^0-E_m^0- i \eta}  |n\rangle \langle m|$ using the regularization of appendix \ref{AppRegularize}. \comm{Note that $\eta$ should always be chosen to be larger than the level spacing if one is interested in systems in the thermodynamic limit.}

\end{widetext}

\section{Unitary driving}\label{SecPeriodic0}
In the case of unitary driving, the projection of the perturbation onto the $P$ space vanishes, $\hat P \cl{L}_1 \hat P=0$, and therefore also 
$\hat P (\cl{L}_0+\cl{L}_1) \hat P=0$. In this case, the formulas fixing the density matrix to zeroth order in the perturbations, Eq.~\eqref{EqUniPert}, and the perturbation theory, presented in Sec.~\ref{SecPertUni} have to be modified compared to the Lindblad dynamics where the inverse of $\hat P \cl{L}_1 \hat P$ within $P$ space is well defined.

\subsection{Projections of the inverse}\label{AppProjInv}
Different projections of full inverse 
$(\hat{P}+\hat{Q})\cl{L}^{-1}(\hat{P}+\hat{Q})$ can, for example, be obtained using the transformations
\begin{align}
\hat{U}
&=\hat{P}+\hat{Q} - (\hat{P}\cl{L}_1 \hat{Q}) \, (\hat{Q}\cl{L}\hat{Q})^{-1} \hat{Q}
\\
\hat{V}
&=\hat{P}+\hat{Q} - \hat{Q} (\hat{Q}\cl{L}\hat{Q})^{-1} \, \hat{Q}\cl{L}_1 \hat{P}\notag
\end{align}
where $\hat{U}$ and $\hat{V}$ are chosen in such a way that they transform $\cl{L}$ into a block-diagonal form \comm{(in $\hat P$ and $\hat Q$ space)},
\begin{align}\label{EqDiagL}
\hat{U}\cl{L} \hat{V}
=\hat{Q}\cl{L}\hat{Q}
-(\hat{P} \cl{L}_1\hat{Q})  (\hat{Q}\cl{L}\hat{Q})^{-1}  (\hat{Q}\cl{L}_1 \hat{P}).
\end{align}
The inverse $\cl{L}^{-1}$ is then obtained from 
\begin{align}
\cl{L}^{-1}
&=
\hat{V} \big( \hat{U}\cl{L}\hat{V} \big)^{-1} \hat{U} \label{EqInv}\\
&=
\hat{V} \big(\hat{P}(\hat{P} \hat{\mathfrak{L}}_2 \hat{P})^{-1}\hat{P} + \hat{Q}(\hat{Q} \cl{L} \hat{Q})^{-1}\hat{Q}\big) \hat{U} \notag
\end{align}
where 
\begin{equation}
\hat{\mathfrak{L}}_2 =-\cl{L}_1\hat{Q} \ (\hat{Q}\cl{L}\hat{Q})^{-1} \ \hat{Q}\cl{L}_1.\end{equation}
Using the definition of $\hat{U}$ and $\hat{V}$ we obtain \comm{for perturbations with $\hat P \cl{L}_1 \hat P=0$} the following expressions for the projections
\begin{align}
\hat{P} \cl{L}^{-1} \hat{P}&=\hat{P}(\hat{P} \hat{\mathfrak{L}}_2\hat{P})^{-1} \hat{P}
\sim \frac{1}{\epsilon^2}  \label{EqPLm1P}\\
\hat{P} \cl{L}^{-1} \hat{Q}&=- \hat{P} (\hat{P} \hat{\mathfrak{L}}_2 \hat{P})^{-1}(\hat{P} \cl{L}_1 \hat{Q}) \ (\hat{Q} \cl{L} \hat{Q})^{-1}\hat{Q} \sim  \frac{1}{\epsilon} \notag \\
\hat{Q} \cl{L}^{-1} \hat{P}&=-\hat{Q}(\hat{Q} \cl{L} \hat{Q})^{-1} \ (\hat{Q} \cl{L}_1 \hat{P}) (\hat{P} \hat{\mathfrak{L}}_2 \hat{P})^{-1} \hat{P} \sim \frac{1}{\epsilon} \notag\\
\hat{Q} \cl{L}^{-1} \hat{Q}&=0 \notag
\end{align}
\comm{$(\hat{P} \hat{\mathfrak{L}}_2\hat{P})^{-1}$ and $(\hat{Q} \cl{L} \hat{Q})^{-1}$ should be interpreted as inverses within $P$ and $Q$ subspace, respectively, while the left hand sides of Eqs.~\eqref{EqPLm1P} correspond to the projections of the full (properly regularized) Liouville inverse onto $P$ and $Q$ subspace.}
Note that the last equation does not imply that $\hat{Q} (\hat{Q} \cl{L} \hat{Q})^{-1}\hat{Q} $ vanishes (which is finite for $\epsilon \to 0$).

A direct consequence of the  equations given above is that in the limit of small $\epsilon$, the inverse in the $P$ sector is given through $\cl{L}_2$ defined in Eq.~\eqref{L2def},
\begin{equation}
\hat{P} \cl{L}^{-1} \hat{P} 
=\hat{P} (\hat{P} \hat{\mathfrak{L}}_2\hat{P})^{-1} \hat{P}
\approx (\hat{P} \cl{L}_2\hat{P})^{-1}\  (1+ \mathcal{O}(\epsilon)).
\end{equation}
 Note that  $\cl{L}_2$ is obtained from $\hat{\mathfrak{L}}_2$ just by replacing  $\hat Q \cl{L} \hat Q$  by $\hat Q \cl{L}_0 \hat Q$ in the inverse.
$\hat{P}\cl{L}_2\hat{P}$ takes over the role of an effective Lindblad super-operator in $P$ space.

\subsection{Zeroth order expansion point}\label{AppUniGGE}
Here we show that the condition \eqref{EqUniPert},
\begin{equation}\label{EqCondUni}
\hat{P}(\cl{L}_1\cl{L}_0^{-1}\cl{L}_1 \rho_0)=0,
\end{equation}
does give the correct reference point $\rho_0$ for situations where $\hat P \cl{L}_1 \hat P=0$.
One way to show this is to use the perturbative analysis provided in Sec.~\ref{SecPertUni} where the power counting of diagrams worked only if the correct reference point was chosen. Below we give a more direct argument.

As described in the main text, Eq.~(\ref{EqCondUni}) is obtained from the requirement that the dominant contribution to the time-averaged expectation value of conserved quantities must vanish,
\begin{equation}\label{aveC}
\ave{\overline{\dot{C}_i}}
=\tr[C_i \hat{P} \cl{L}_1 (\rho_0+\delta\rho)]
=\tr[C_i \hat{P} \cl{L}_1 \delta\rho]	
\stackrel{!}{=}0,
\end{equation}
where $\hat{P}$ is used to extract the non-oscillatory component and $\tr[C_i \hat{P} \cl{L}_1 \rho_0]=0$ due to the cyclicity of trace.
As discussed in the main text, the starting point is the exact formula for $\delta \rho$, Eq.~\eqref{EqDeltaRho} and the formula Eq.~(\ref{EqXY}) which directly leads to
\begin{align}
\delta\rho
&=\delta\rho^{(I)}+\delta\rho^{(II)}\\
&=-\cl{L}_0^{-1} \cl{L}_1\rho_0 + \cl{L}^{-1}\cl{L}_1\cl{L}_0^{-1}\cl{L}_1\rho_0.\notag
\end{align}
If we use only $\delta\rho^{(I)}$ in Eq.~\eqref{aveC}, then Eq.~\eqref{EqCondUni} follows immediately.  Equivalently, 
the condition Eq.~\eqref{EqCondUni} implies that the contribution from  $\delta\rho^{(I)}$ vanishes in Eq.~\eqref{aveC}.
In the following we will show that Eq.~\eqref{EqCondUni} also implies that the contribution from $\delta\rho^{(II)}$ to Eq.~\eqref{aveC} vanishes, which is less obvious, and a useful consistency check.

Plugging $\delta\rho^{(II)}$ into Eq.~\eqref{aveC} one finds
\begin{align}
\tr[& C_i \hat{P} \cl{L}_1 \delta\rho^{(II)}] \notag \\
&=\tr[C_i \hat{P} \cl{L}_1   \cl{L}^{-1}  \cl{L}_1 \cl{L}_0^{-1}\cl{L}_1\rho_0] \notag \\
&=\tr[C_i \hat{P} \cl{L}_1  \hat{Q}  \cl{L}^{-1} \hat{Q} \cl{L}_1 \cl{L}_0^{-1}\cl{L}_1\rho_0]=0
\end{align}
where the third line differs from the first one by two extra $\hat Q$ super-operators enclosing $ \cl{L}^{-1} $. The first one can be inserted because we consider the case $\hat{P} \cl{L}_1 \hat P=0$ and therefore $\hat{P} \cl{L}_1 =\hat{P} \cl{L}_1  \hat Q$. The second one can be used as
a consequence of  Eq.~\eqref{EqCondUni}, which states that the $\hat P$ projection of the operator to the right of $\cl{L}^{-1}$ vanishes. Finally, we can use that $\hat{Q} \cl{L}^{-1} \hat{Q}=0$, see Eq.~\eqref{EqPLm1P}, to prove that the whole expression vanishes. 

To finish our argument, we still have to show that $\delta \rho$ is small for $\epsilon \to 0$ provided that Eq.~\eqref{EqPLm1P} holds, which can be done using similar arguments as above.  First, the combination $\cl{L}_0^{-1} \cl{L}_1\rho_0=\cl{L}_0^{-1} \hat Q \cl{L}_1\rho_0$ is non-singular for $\hat{P} \cl{L}_1 \hat P=0$ which implies that $\delta\rho^{(I)}\sim \mathcal O(\epsilon)$. 
Second, we used already above that $\delta\rho^{(II)} =\cl{L}^{-1}\cl{L}_1\cl{L}_0^{-1}\cl{L}_1\rho_0=\cl{L}^{-1}\hat Q\cl{L}_1\cl{L}_0^{-1}\cl{L}_1\rho_0$. As $\hat Q \cl{L}^{-1}\hat Q=0$ and $\hat P \cl{L}^{-1}\hat Q\sim  \mathcal O(1/\epsilon)$ it follows immediately that  also $\delta\rho^{(II)}\sim \mathcal O(\epsilon)$ which concludes the derivation of
Eq.~\eqref{EqCondUni}.

\subsection{Monochromatic driving }\label{monochromatic}
Some of the diagrams depicted in Fig.~\ref{figDiagramsNew} vanish when a perturbation which contains only oscillations with a single frequency $\omega$ is considered. In this case $\hat{P}(\cl{L}_1)^3\hat{P}=0$ since each application of $\cl{L}_1$ changes the Floquet index $n$ of $\rho^{(n)}$, Eq.~\eqref{EqFloquet}, by $\pm 1$. Consequently, for example, the $\mathcal{O}(\epsilon)$ diagram in the $(\parallel)$ sector vanishes and conservation laws are only changed by processes of order $\mathcal{O}(\epsilon^2)$, represented by the diagram in the second line of Fig.~\ref{figDiagramsNew}  (note that the third one vanishes as well). 
The situation is different in the presence of driving with higher harmonics, e.g. $\cos(\omega t)$ and $\cos(2\omega t)$, when $\hat{P}(\cl{L}_1)^3\hat{P}\neq 0$. In this case the dominant stationary correction to the expectation value of $C_i$ is of $\mathcal{O}(\epsilon)$, as shown by the diagram in the first line in the $(\parallel)$ sector.



\end{document}